\documentclass[12pt]{article}%
\usepackage{amsmath}
\usepackage{amsfonts}
\usepackage{amssymb}
\usepackage{graphicx, epsfig}
\usepackage{graphicx}%
\providecommand{\U}[1]{\protect\rule{.1in}{.1in}}
\providecommand{\U}[1]{\protect\rule{.1in}{.1in}}
\newtheorem{theorem}{Theorem}

\newenvironment{changemargin}[2]{
\begin{list}{}{\setlength{\topsep}{0pt}\setlength{\leftmargin}{0pt}\setlength{\rightmargin}{0pt}\setlength{\listparindent}{\parindent}
\setlength{\itemindent}{\parindent}\setlength{\parsep}{0pt plus 1pt}\addtolength{\leftmargin}{#1}\addtolength{\rightmargin}{#2}}\item }
{\end{list}}
\begin{document}

\title{Robust estimators for generalized linear models with a dispersion parameter}
\author{Michael Amiguet$^{(1)}$, Alfio Marazzi$^{(1)}$, Marina Valdora$^{(2)} $
\and and Victor Yohai$^{(2)}$,\ \\{\small (1) Universit\'{e} de Lausanne, (2) Universidad de Buenos Aires}}
\maketitle

\begin{abstract}
Highly robust and efficient estimators for the generalized linear model with a
dispersion parameter are proposed. The estimators are based on three steps. In
the first step the maximum rank correlation estimator is used to consistently
estimate the slopes up to a scale factor. In the second step, the scale
factor, the intercept, and the dispersion parameter are consistently estimated
using a MT-estimator of a simple regression model. The combined estimator is
highly robust but inefficient. Then, randomized quantile residuals based on
the initial estimators are used to detect outliers to be rejected and to
define a set S of observations to be retained. Finally, a conditional maximum
likelihood (CML) estimator given the observations in S is computed. We show
that, under the model, S tends to the complete sample for increasing sample
size. Therefore, the CML tends to the unconditional maximum likelihood
estimator. It is therefore highly efficient, while maintaining the high degree
of robustness of the initial estimator. The case of the negative binomial
regression model is studied in detail.

\end{abstract}

\section{Introduction\label{introduction}}

In recent years, several extensions of the generalized linear models (GLM;
Nelder and Wedderburn, 1972) have been proposed to increase flexibility in
modelling complex data structures. We consider the case where the response
distribution does not necessarily belong to the exponential family and where a
dispersion parameter is present. For this case, we will propose highly
efficient and highly robust estimators. We focus on the Negative Binomial (NB)
regression model, but we also consider the Beta regression model as an example
with continuous response. NB regression (see Hilbe, 2008) extends Poisson
regression for modeling count data in presence of overdispersion. Beta
regression (Ferrari and Cribari-Neto, 2004) is a tool for modelling continuous
responses which are restricted to the interval $[0,1]$, such as rates and
proportions. Both these models have important biometrical applications. NB
regression is the most popular model for the analysis of hospital length of
stay (e.g., Austin et al., 2002; Hilbe, 2008; Carter and Potts, 2014). Among
other applications, we also mention its use to model falls data (Aeberhard et
al., 2014). Applications of Beta regression in medicine can be found in Hunger
et al. (2011), Swearingen et al. (2011), and Seow et al. (2012) among others.

Usually, the parameters are estimated by means of the maximum likelihood (ML)
principle, which provides fully efficient estimators when the observations
follow the model.\ ML procedures to fit the NB regression have been
implemented in popular statistical software such as STATA, SAS, SPSS, and in
the R package MASS (Venables and Ripley, 1999). An implementation of the Beta
regressions can be found in R (Cribari--Neto and Zeiles, 2010).

Unfortunately, the ML estimator is extremely sensitive to the presence of
outliers in the sample, i.e., observations with unexpectedly extreme values in
the response variable. This sensitivity increases when these extreme responses
come together with large values in the covariate space. In certain
applications, such as the analysis of hospital length of stay, the proportion
of outliers - often called the contamination level - may be as high as $10\%$.
Such a level of contamination can not only strongly bias the coefficient
estimates but also lead to overestimating the dispersion parameter. As a
consequence, inferences based on the ML fit may be badly misguided.

There are two basic approaches to detect outliers and assess their influence.
The first one makes use of diagnostic tools based on ML residuals. Specific
proposals for GLM are described by Davison and Snell (1991) and proposals for
Beta regression by Espinheira et al. (2008) and Rocha and Simas (2010).
However, this strategy may fail because the ML estimators may be distorted
and residuals corresponding to outliers are not necessarily large and visible;
a well known \textquotedblleft masking effect\textquotedblright\ is described
in Maronna et al. (2006, p. 179). A better strategy, is the use of a robust
estimator, that is an estimator which is not very sensitive to the presence of
outliers. There are many proposals of robust estimators for GLM models (e.g.,
K\"{u}nsch et al. 1989; Cantoni and Ronchetti, 2001). However, most of them do
not admit an extra parameter besides the coefficient vector. A few robust
estimators of the parameters of the NB distribution in the absence of
covariates have been considered in Cadigan and Chen (2001) and Amiguet (2011).
Marazzi and Yohai (2010) implemented M estimators satisfying Hampel's
optimality principle (Hampel et al., 1986) for multiparameter families of
distributions including NB and Beta. Yet, it is cumbersome to extend these
estimators to the regression case. Aeberhard et al. (2014) proposed a
generalized M (GM) estimator for NB regression. Unfortunately, GM estimators
have several drawbacks. In particular, their degree of robustness - as
measured by the breakdown point - decreases when the number of covariates
increases (Maronna et al., 2006, p.149). Moreover, GM estimators depend on
\textquotedblleft tuning constants\textquotedblright\ that are chosen to
attain a given level of efficiency at a specified model with known parameter
values; but the parameters are unknown before estimation. In order to ensure
consistency at the (unknown) model, several corrections have to implemented
adding complexity to the computation and increasing the computing time. At
present, no robust procedure for Beta regression has been published.

In this paper, we introduce a novel class of estimators for GLMs with a
dispersion parameter. Following an approach that we have developed in previous
papers for different models (Marazzi and Yohai, 2004; Locatelli, Marazzi,
Yohai, 2010), we consider a three phase procedure. In the first phase, a
highly robust but possibly inefficient estimator is computed. This initial
estimator allows outlier identification. Finally, a conditional ML procedure
is used, where observations are constrained to belong to a subsample free of
large outliers. However, in the absence of outliers, this subsample tends to
the original sample if its size increases and, therefore, the final estimator
is asymptotically fully efficient. Nevertheless, it maintains a similar degree
of robustness as the initial estimator. Conditional ML estimators have also
been used by Cuesta-Albertos, Matr\'{a}n, and Mayo-Iscar (2008) to define
multivariate robust location and dispersion estimators.

In Section \ref{The model} we introduce the general model. Section
\ref{Est proc} defines the estimators. The efficiency and the robustness of
the new procedures are demonstrated in Section \ref{simulations} by means of
Monte Carlo experiments. Two examples, where the procedures are applied to
hospital length of stay data are described in Section \ref{Examples}. The
discussion in Section \ref{Discussion} ends the paper. Three appendices
provide proofs and some supplementary material. The methods we are proposing
in this paper have been implemented in the R package \textquotedblleft
robustGLM\textquotedblright\ available on the Comprehensive R Archive
Network.\medskip

\section{\textbf{The model\label{The model}}}

Let $F_{\mu,\alpha}(y)$ denote a general family of discrete or continuous
distribution functions, where $\mu$ is the mean and $\alpha$ is a
dispersion\ parameter, and let $f_{\mu,\alpha}(y)$ denote the corresponding
probability (density) function. We will focus on two specific examples of
families, one discrete and one continuous :\smallskip

\noindent- the NB family:%
\begin{equation}
f_{\mu,\alpha}(y)=\frac{\Gamma(y+1/\alpha)}{\Gamma(1/\alpha)\Gamma
(y+1)}(\alpha\mu+1)^{-1/\alpha}\left(  \frac{\alpha\mu}{\alpha\mu+1}\right)
^{y}\text{, }y=0,1,2,...\text{, }\alpha\geq0\text{, }\mu\geq0\text{;}
\label{fNB}%
\end{equation}
\noindent- the Beta family:%
\begin{equation}
f_{\mu,\alpha}(y)=\frac{\Gamma(1/\alpha)}{\Gamma(\mu/\alpha)\Gamma
((1-\mu)/\alpha)}y^{\mu/\alpha-1}(1-y)^{(1-\mu)/\alpha-1}\text{, }0<y<1\text{,
}\alpha\geq0\text{, }\mu\geq0\text{.} \label{fBeta}%
\end{equation}

\noindent In both cases, the parametrization has been chosen so that the
expected value is $\mu$. In the NB case, the variance is $\mu+\alpha\mu^{2}$;
in the Beta case, the variance is $\mu(1-\mu)/(1+1/\alpha)$. In both cases,
fixing $\mu$, the variance increases with $\alpha$.

We will need the following assumption on $F_{\mu,\alpha}(y)$, which is
satisfied in our examples:\smallskip

\noindent\textsl{Assumption A}\emph{:} For any $\alpha$, $Y_{1}\sim F_{\mu
_{1},\alpha}(y)$, $Y_{2}\sim F_{\mu_{2},\alpha}(y)$, if $\mu_{2}>\mu_{1} $
then $Y_{2}$ $\succ Y_{1}$, where \textquotedblleft$\succ$\textquotedblright%
\ means \textquotedblleft stochastically larger\textquotedblright.\smallskip

Suppose that a response $Y$ and a vector $\mathbf{X}=(X_{1},...X_{p}%
)^{\text{T}}$ of covariates are observed. We consider the following class of
regression models%
\begin{equation}
Y\text{ }|\text{ }\mathbf{X=x}\sim F_{h(\mathbf{x}^{\text{T}}\mathbf{\beta
}_{0}),\alpha_{0}}\text{,} \label{model}%
\end{equation}
where $h$ is a strictly increasing known link function, and $\mathbf{\beta
}_{0}=(\beta_{01},...\beta_{0p})^{\text{T}}$ is a vector of coefficients. We
assume that $X_{1}$ is constantly equal to one, that is, $\beta_{01}$ is an
intercept. We will use the notations $\mathbf{x}^{\text{T}}=(1,\mathbf{x}%
^{\ast\text{T}})$, $\mathbf{\beta}_{0}^{\text{T}}=(\beta_{01},\mathbf{\beta
}_{0}^{\ast\text{T}})$, $\mathbf{\gamma}_{0}=\mathbf{\beta}_{0}^{\ast
}/||\mathbf{\beta}_{0}^{\ast}||$, and $\mathbf{\theta}=(\mathbf{\beta}%
,\alpha)$.

We assume that a random sample $(\mathbf{x}_{1},y_{1})\,,...,(\mathbf{x}%
_{n},y_{n})$ is available. The ML estimator of $\mathbf{\theta}_{0}=\left(
\mathbf{\beta}_{0},\alpha_{0}\right)  $ maximizes the log-likelihood of the
sample given by
\[
\mathcal{L(}\mathbf{\theta})=\sum_{i=1}^{n}\ln\left(  f_{h(\mathbf{x}%
_{i}^{\text{T}}\mathbf{\beta}),\alpha}(y_{i})\right)  \text{.}%
\]
The ML estimator is very efficient but not robust. We want to obtain highly
robust and efficient estimators of $\mathbf{\beta}_{0}$ and $\alpha_{0}%
$.\medskip

\section{ \textbf{Estimation procedure}\label{Est proc}}

The proposed procedure starts with the computation of a very robust but not
necessarily efficient initial estimator which provides the tool for outlier
identification. Then, a conditional ML approach is used - where the outliers
are removed - which provides a fully efficient estimator.

Most familiar highly robust estimators of regression, such as LMS, LTS, and S
estimators (see, e.g., Maronna et al., 2006), are based on the minimization of
a robust measure of the residual scale, such as an M scale (Huber, 1980).
These estimators have been used as initial estimators of well known highly
robust and efficient procedures, such as MM (Yohai, 1987), and TML (Marazzi
and Yohai, 2004) estimators. However, for the regression models we are
considering here, a different approach has to be used because the residual
distribution may depend on the covariates and residual measures of scale are
not available in this case. We therefore propose an approach based on the
maximum rank correlation (MRC) estimator introduced by Han (1987a) and Han
(1987b). However, the MRC estimator identifies the scaled slopes
$\mathbf{\gamma}_{0}=\mathbf{\beta}_{0}^{\ast}/||\mathbf{\beta}_{0}^{\ast}||$,
but it does not identify the intercept $\beta_{01}$, the dispersion parameter
$\alpha_{0}$, and the scale factor $\eta_{0}=||\mathbf{\beta}_{0}^{\ast}||$.
So, we need to estimate these three parameters separately. The complete
proposal can then be summarized as follows:\smallskip

\begin{description}
\item[Step 1] Compute the MRC estimator $\mathbf{\tilde{\gamma}}$ of
$\mathbf{\gamma}_{0}$. In addition, compute robust and consistent estimators
$\tilde{\beta}_{1}$, $\tilde{\alpha}$, and $\tilde{\eta}$, of $\beta_{01}$,
$\alpha_{0}$, and $\eta_{0}$. Then, initial estimators of $\mathbf{\beta}_{0}$
and $\alpha_{0}$ are given by $\mathbf{\tilde{\beta}}=(\tilde{\beta}%
_{1},\tilde{\eta}\mathbf{\tilde{\gamma}}^{\text{T}})^{\text{T}}$ and
$\tilde{\alpha}$ respectively.

\item[Step 2] Compute randomized quantile residuals $z_{i}$ (Dunn and Smyth,
1996) based on the initial model and use them to define cutoff values
$\tilde{a}$ and $\tilde{b}$, so that influential outliers are defined as
observations such that $z_{i}\notin\lbrack\tilde{a},\tilde{b}]$.

\item[Step 3] Compute a conditional ML estimator of $\mathbf{\theta}_{0}$
given $z_{i}\in\lbrack\tilde{a},\tilde{b}]$.\smallskip
\end{description}

\noindent In the following subsections, we provide a detailed description of
each single step.\medskip\ 

\subsection{The initial estimator\textit{\label{inest}}}

For a given coefficient vector $\mathbf{\gamma}=(\gamma_{2},...,\gamma
_{p})^{\text{T}}$, the Kendall's $\tau$ correlation coefficient between the
responses $y_{i}$-s and the linear combinations $\mathbf{\gamma}^{\text{T}%
}\mathbf{x}_{i}^{\ast}$-s is given by
\[
\tau(\mathbf{\gamma)=}\frac{1}{n(n-1)}\sum_{i\neq j}I\left[  (\mathbf{\gamma
}^{\text{T}}\mathbf{x}_{j}^{\ast}-\mathbf{\gamma}^{\text{T}}\mathbf{x}%
_{i}^{\ast})(y_{j}-y_{i})\geq0\right]
\]
and the maximum rank correlation (\textit{MRC) estimator} of $\mathbf{\gamma
}_{0}$ is defined by
\begin{equation}
\mathbf{\tilde{\gamma}}=\arg\min_{\left\Vert \mathbf{\gamma}\right\Vert
=1}\tau(\mathbf{\gamma})\text{.} \label{MReq}%
\end{equation}
\smallskip The robustness of Kendall's $\tau$ correlation coefficient has been
studied by Alfons et al. (2016). Under the assumption A, the MRC estimator
strongly converges to $\mathbf{\gamma}_{0}$ for any strictly increasing $h$
(Han, 1987a); it is also root $n$ consistent and asymptotically normal
(Sherman, 1993)\textsl{. }

To compute the MRC estimator one can utilize a subsampling procedure. Note
that the simple evaluation of the objective function requires $O(n^{2})$
calculations, but an algorithm using $O(n$ $\log$ $n)$ calculations has been
proposed by Abrevaya (1999). However, in the Monte Carlo experiments described
in Section \ref{simulations}, we used the very fast function maxCorGrid of the
R package ccaPP (Alfons, 2015) based on an alternate grid algorithm described
in Alfons et al. (2016).

We now turn to the estimation of $\beta_{01}$, $\alpha_{0}$, and $\eta_{0}
$,\ necessary to complete the initial estimator. We observe that
$h(\mathbf{x}^{\text{T}}\mathbf{\beta}_{0})=h(\beta_{01}+\eta_{0}%
\mathbf{\gamma}_{0}^{\text{T}}\mathbf{x}^{\ast})$. Since $\mathbf{\tilde
{\gamma}}$ is close to $\mathbf{\gamma}_{0}$, we approximate $\mathbf{\gamma
}_{0}^{\text{T}}\mathbf{x}_{i}^{\ast}$ by $v_{i}=\mathbf{\tilde{\gamma}%
}^{\text{T}}\mathbf{x}_{i}^{\ast}$\ and consider the simple regression model
with just one covariate:\textsl{\ }%
\begin{equation}
Y\text{ }|\text{ }v\sim F_{h(\beta_{01}+\eta_{0}\nu),\alpha_{0}}\text{.}
\label{simplemodel}%
\end{equation}
For this model and a given value $\alpha$ of the unknown $\alpha_{0}$, we have
many highly robust estimators $\tilde{\beta}_{1}^{\ast}(\alpha)$, $\tilde
{\eta}^{\ast}(\alpha)$, of $\beta_{01}$ and $\eta_{0}$. Examples are: the
conditionally unbiased bounded influence estimator of K\"{u}nsch et al.
(1989), the RQL estimator of Cantoni and Ronchetti (2001), and the weighted MT
estimators of Valdora and Yohai (2014)\textsl{. }Finally, to estimate
$\alpha_{0}$, we consider a bounded function $\psi(y,%
%TCIMACRO{\U{3bc} }%
%BeginExpansion
\mu
%EndExpansion
,\alpha)$ such that, for all $%
%TCIMACRO{\U{3bc} }%
%BeginExpansion
\mu
%EndExpansion
$, we have%
\begin{equation}
E_{\mu,\alpha}\left[  \psi(y,%
%TCIMACRO{\U{3bc} }%
%BeginExpansion
\mu
%EndExpansion
,\alpha)\right]  =0. \label{Epsi}%
\end{equation}
Then, for any fixed $\mu,$ if $y_{1},y_{2},...y_{n\text{ }}$is a random sample
of $NB(\mu,\alpha_{0})$, the M estimator of $\alpha$ satisfying the equation%
\[
\sum_{i=1}^{n}\psi(y_{i},%
%TCIMACRO{\U{3bc} }%
%BeginExpansion
\mu
%EndExpansion
,\alpha)=0
\]
is Fisher consistent for $\alpha_{0}$. Then, an initial consistent estimator
$\tilde{\alpha}$ of $\alpha_{0}$ is obtained by solving%
\begin{equation}
\sum_{i}\psi(y_{i},h(\tilde{\beta}_{1}^{\ast}(\alpha)+\tilde{\eta}^{\ast
}(\alpha)v_{i}),\alpha)=0\text{.} \label{MEalpha1}%
\end{equation}
The Fisher consistency of this estimator is immediate. In fact,
asymptotically, $h(\tilde{\beta}_{1}^{\ast}(\alpha)+\tilde{\eta}^{\ast}%
(\alpha)v_{i})=E(y_{i})$, and then by (\ref{Epsi})%
\[
E(\psi(y,\mu_{i},\alpha))=0.
\]
Once $\tilde{\alpha}$ is computed, we define the initial estimators of
$\beta_{01}$ and $\eta_{0}$ by $\tilde{\beta}_{1}=\tilde{\beta}_{1}^{\ast
}(\tilde{\alpha})$, $\tilde{\eta}=\tilde{\eta}^{\ast}(\tilde{\alpha})$. In
this way we obtain the initial estimators $\mathbf{\tilde{\beta}}%
=(\tilde{\beta}_{1},\tilde{\eta}\mathbf{\tilde{\gamma}})$ of $\mathbf{\beta
}_{0}$ and $\tilde{\alpha}$ of $\alpha_{0}$.

We will assume that:\smallskip

\noindent\textsl{Assumption B}: $n^{1/2}(\mathbf{\tilde{\beta}-\beta}%
_{0})=O_{p}(1)$ and $n^{1/2}(\tilde{\alpha}-\alpha_{0})=O_{p}(1).\smallskip$

In the simulations of Section \ref{simulations} and the examples in Section
\ref{Examples}, we use a weighted MT estimator for $\tilde{\beta}_{1}^{\ast
}(\alpha)$, $\tilde{\eta}^{\ast}(\alpha)$ (see appendix \ref{EMT}) and the
score function $\psi$ of the optimal bounded influence estimator according to
Hampel (1972) described in Marazzi and Yohai (2010). It can be proved that,
under general conditions, the resulting initial estimators $\mathbf{\tilde
{\beta}}$ and $\tilde{\alpha}$ satisfy the assumption B.

\subsection{Adaptive cutoff values and outlier detection\label{ACF}}

We now assume that some preliminary estimator $\mathbf{\tilde{\theta}%
}=(\mathbf{\tilde{\beta}},\tilde{\alpha})$ of $\mathbf{\theta}_{0}$ is
available, for example the estimators defined in the previous section. Since
the residual distribution depends on the covariates, residuals cannot be used
in the usual way for the purpose of highlighting outliers. Instead, we use the
randomized quantile residuals\ (RQR) that were proposed in Dunn and Smyth
(1996) for exploratory purposes. Let $\tilde{\mu}_{\mathbf{x}}=h(\mathbf{x}%
^{\text{T}}\mathbf{\tilde{\beta}})$. Then, the RQRs are defined by
\[
z_{i}=F_{\tilde{\mu}_{\mathbf{x}},\tilde{\alpha}}(y_{i})
\]
in the continuous case and by
\[
z_{i}=F_{\tilde{\mu}_{\mathbf{x}},\tilde{\alpha}}(y_{i})-u_{i}f_{\tilde{\mu
}_{\mathbf{x}},\widetilde{\alpha}}(y_{i})
\]
\ in the discrete case, where $\left\{  u_{1},...,u_{n}\right\}  $ is a sample
from a uniform distribution $U[0,1]$ independent of the original sample
$(\mathbf{x}_{1},y_{1})\,,...,(\mathbf{x}_{n},y_{n})$.

If $\mathbf{\tilde{\theta}}=\mathbf{\theta}_{0}$, $\left\{  z_{1}%
,...,z_{n}\right\}  $ is a sample from $U[0,1]$. Then, a fixed lower cutoff
value $a$ and a fixed upper cutoff value $b$ for the RQRs are simply given by
a low, respectively a large quantile of $U[0,1]$ -- e.g., $a=0.05$ and
$b=0.95$ -- and observations such that $z_{i}\notin\lbrack a,b]$ may be
identified as outliers.\ However, we propose the use of \textquotedblleft
adaptive\textquotedblright\ cutoff values $\tilde{a}$ and $\tilde{b}$ that,
under the assumed model, tend to $0$ and $1$ respectively, when
$\mathbf{\tilde{\beta}}$ and $\tilde{\alpha}$ are consistent estimators.
Therefore, under the model, i.e., in the absence of outliers, the fraction of
observations that are erroneously identified as outliers tends to $0$ when the
sample size $n\rightarrow\infty$.

To define the adaptive cutoff values, we follow a procedure similar to the
ones described in Marazzi and Yohai (2004, Section 3.2) and in Gervini and
Yohai (2002). Let $F_{n}$ denote the empirical cdf of $z_{1},...,z_{n}$ and
$F_{n,t}^{R}$ and $F_{n,t}^{L}$ be the right and the left truncated versions
of $F_{n}$ for a given $t$ respectively, i.e,.
\[
F_{n,t}^{R}(z)=\left\{
\begin{array}
[c]{cc}%
F_{n}(z)/F_{n}(t) & \text{if }z\leq t,\\
1 & \text{otherwise,}%
\end{array}
\right.
\]%
\[
F_{n,t}^{L}(z)=\left\{
\begin{array}
[c]{cc}%
(F_{n}(z)-F_{n}(t))/(1-F_{n}(t)) & \text{if }z\geq t,\\
0 & \text{otherwise.}%
\end{array}
\right.
\]
We then compare the rights tails of $F_{n}$ and the $U[0,1]$, looking for the
largest $t$ such that $F_{n,t}^{R}(z)\geq z$ for all $z\geq\zeta_{2}$ where
$\zeta_{2}$ is a value close to one. More precisely, we define an upper cutoff
value as
\[
\tilde{b}=\sup\left\{  t:\inf_{z\geq\zeta_{2}}(F_{n,t}^{R}(z)-z)\geq0\right\}
.
\]
In a similar way, we define a lower cutoff value as
\[
\tilde{a}=\inf\left\{  t:\sup_{z\leq\zeta_{1}}(F_{n,t}^{L}(z)-z)\leq0\right\}
,
\]
where $\zeta_{1}$ is close to zero.

We assume that:\smallskip

\noindent\textsl{Assumption C}: The density $f(y,\mu,\alpha)$ has a bounded
derivative with respect to $\mu$ and $\alpha$.$\smallskip$

\noindent Then, we have the following Theorem, proved in Appendix
\ref{proofT1}.

\begin{theorem}
\label{conv a b}Assume B and C. Then
\[
n^{1/2}\tilde{a}=O_{p}(1),\text{ \ }n^{1/2}(\tilde{b}-1)=O_{p}(1).
\]

\end{theorem}

Usually, a quite high value of is $\zeta_{2}$ chosen. Our usual choice is
$\zeta_{2}=0.95$; however, in the presence of a large proportion of high
outliers, it may be convenient to use a lower value, e.g., $\zeta_{2}=0.90$.
Similar considerations apply to the choice of the lower cutoff $\tilde{a}$ and
we usually set $\zeta_{1}=0.05$, but $\zeta_{1}=0.10$ would allow removing a
larger fraction of small observations, such as \textquotedblleft excess
zeros\textquotedblright in the NB case. (In fact, a very small $\zeta_{1}$
could fail to identify many \textquotedblleft excess zeros\textquotedblright,
because each one of them corresponds to several distinct $z_{i}$'s and may not
emerge as an extremely small value.)\medskip\ 

\subsection{Final estimator\label{Final est}}

In the final step, we improve the efficiency of the initial estimator using a
conditional ML approach. Suppose first that fixed cutoff values $a$ and
$b\ $are given and the RQRs are computed. Let $p_{\mathbf{\beta,}\alpha}%
^{\ }(y$ $|$ $\mathbf{x,}$ $Z\mathbf{\in}[a,b])$ denote the conditional
density of $Y$ given $\mathbf{X=x}$ and $Z\in\lbrack a,b]$, where $Z\sim
U[0,1]$ represents the RQR. Then, the conditional density is of the form
\begin{equation}
p_{\mathbf{\beta,}\alpha}^{\ }(y\text{ }|\text{ }\mathbf{x,}\text{
}Z\mathbf{\in}[a,b])=f_{h(\mathbf{x}_{i}^{\text{T}}\mathbf{\beta}),\alpha
\ }(y)W_{a,b}(\mathbf{x},\mathbf{\beta,}\alpha)\text{.} \label{conddistr}%
\end{equation}
In the continuous case we have
\[
W_{a,b}(\mathbf{x},\mathbf{\beta,}\alpha)=\frac{I(F_{\mu_{\mathbf{x}},\alpha
}^{-1}(a)\leq\ y\leq F_{\mu_{\mathbf{x}},\alpha}^{-1}(b))}{b-a}.
\]
In the discrete case, the following expression (\ref{zWeights}) for
$W_{a.b}(\mathbf{x},\mathbf{\beta,}\alpha)$ is derived in the appendix
\ref{proofzw}. Let, for any $c$,
\[
y_{\mathbf{x}\ }^{\ast}(c)=\max\{y:F_{\mu_{\mathbf{x}},\alpha}(y)\leq c\},
\]
and
\[
t_{c,\mathbf{x}}=\frac{F_{\mu_{\mathbf{x}},\alpha}(y_{\mathbf{x}}^{\ast
}(c)+1)-c}{f_{\mu_{\mathbf{x}},\alpha}(y_{\mathbf{x}}^{\ast}(c)+1)},
\]
Put
\[
T_{a,\mathbf{x}}=y_{\mathbf{x}}^{\ast}(a)+2\text{, \ }T_{b,\mathbf{x}%
}=y_{\mathbf{x}}^{\ast}(b),
\]%
\[
A_{\mathbf{x}}=\{y:T_{a,\mathbf{x}}\leq y\leq T_{b,\mathbf{x}}\},
\]
and%
\[
Q(\mathbf{x,\beta,}\alpha)=F_{\mu_{\mathbf{x}},\alpha}(T_{b,\mathbf{x}%
})-F_{\mu_{\mathbf{x}},\alpha}(T_{a,\mathbf{x}}-1)+f_{\mu_{\mathbf{x}},\alpha
}(T_{a,\mathbf{x}}-1)t_{a,\mathbf{x}}+f_{\mu_{\mathbf{x}},\alpha
}(T_{b,\mathbf{x}}+1)(1-t_{b,\mathbf{x}}).
\]
Then
\begin{equation}
W_{a,b}(\mathbf{x},\mathbf{\beta,}\alpha)=\left\{
\begin{array}
[c]{ccc}%
\frac{1}{Q(\mathbf{x,\beta,}\alpha)} & \text{if} & y\in A_{\mathbf{x}},\\
\frac{t_{a,\mathbf{x}}}{Q(\mathbf{x,\beta,}\alpha)} & \text{if} &
y=T_{a,\mathbf{x}}-1,\\
\frac{1-t_{b,\mathbf{x}}}{Q(\mathbf{x,\beta,}\alpha)} & \text{if} &
y=T_{b,\mathbf{x}}+1,\\
0 & \text{if} & \text{elsewhere.}%
\end{array}
\right.  \ \label{zWeights}%
\end{equation}

We now suppose that $\tilde{a}$ and $\tilde{b}$ are the adaptive cutoff values
defined above, and consider the adaptive conditional likelihood function
\[
\mathcal{L}_{\text{CML}}\ \mathcal{(}\mathbf{\theta})=\sum_{\ i=1}%
^{\ n}I(\tilde{a}\leq z_{i}\leq\tilde{b})\ln\left(  p_{\mathbf{\beta,}\alpha
}^{\ }(y_{i}\text{ }|\text{ }\mathbf{x}_{i}\mathbf{,}\text{ }z_{i}\mathbf{\in
}[\tilde{a},\tilde{b}])\right)  .
\]
The \textit{conditional maximum likelihood} (\textit{CML) estimator
}$\mathbf{\hat{\theta}}_{\text{CML}}=(\mathbf{\hat{\beta}}_{\text{CML}}%
,\hat{\alpha}_{\text{CML}})$ is defined by
\[
\mathbf{\hat{\theta}}_{\text{CML}}=\arg\max_{\mathbf{\theta}}\mathcal{L}%
_{\text{CML}\ }\mathcal{(}\mathbf{\theta}).
\]
In the discrete case, a slight modification of this definition is convenient.
We note that (see appendix \ref{proofzw}):
\[
\{a\leq z_{i}\leq b\}=\{T_{a,\mathbf{x}_{i}}\leq y_{i}\leq T_{b,\mathbf{x}%
_{i}}\}\ \cup\{y_{i}=T_{l,\mathbf{x}}^{\ast},\text{ }u_{i}\leq t_{a,\mathbf{x}%
_{i}}\}\cup\{y_{i}=T_{u,\mathbf{x}}^{\ast},\text{ }u_{i}\geq t_{b,\mathbf{x}%
_{i}}\},
\]
where $T_{l,\mathbf{x}}^{\ast}=T_{a,\mathbf{x}}-1,$ and $T_{u,\mathbf{x}%
}^{\ast}=T_{b,\mathbf{x}}+1$. Then,
\begin{align*}
\mathcal{L}_{\text{CML}}\mathcal{(}\mathbf{\theta})=  &  \sum_{T_{\tilde
{a},\mathbf{x}_{i}}\leq y_{i}\leq T_{\tilde{b},\mathbf{x}_{i}}\ }^{\ }%
\ln\left(  p_{\mathbf{\beta,}\alpha}^{\ }(y_{i}\text{ }|\text{ }\mathbf{x}%
_{i}\mathbf{,}\text{ }z_{i}\mathbf{\in}[\tilde{a},\tilde{b}])\right) \\
&  +\sum_{y_{i}=T_{l,\mathbf{x}}^{\ast}}^{\ }I(u_{i}\leq t_{\tilde
{a},\mathbf{x}_{i}})\ln\left(  p_{\mathbf{\beta,}\alpha}^{\ }(y_{i}\text{
}|\mathbf{x}_{i}\mathbf{,}\text{ }z_{i}\mathbf{\in}[\tilde{a},\tilde
{b}])\right) \\
&  +\sum_{y_{i}=T_{u,\mathbf{x}}^{\ast}}^{\ }I(u_{i}\geq t_{\tilde
{b},\mathbf{x}_{i}})\ln\left(  p_{\mathbf{\beta,}\alpha}^{\ }(y_{i}\text{
}|\mathbf{x}_{i}\mathbf{,}\text{ }z_{i}\mathbf{\in}[\tilde{a},\tilde
{b}])\right)  .
\end{align*}
Since \ the $u_{i}$-s are non--informative, we replace $I(u_{i}\leq
t_{\tilde{a},\mathbf{x}_{i}})$ and $I(u_{i}\geq t_{\tilde{b},\mathbf{x}_{i}})$
by their expected values, and define%
\begin{align*}
\mathcal{L}_{\text{MCML}}\mathcal{(}\mathbf{\theta})  &  =\sum_{T_{\tilde
{a},\mathbf{x}_{i}}\leq y_{i}\leq T_{\tilde{b},\mathbf{x}_{i}}\ }^{\ }%
\ln\left(  p_{\mathbf{\beta,}\alpha}^{\ }(y_{i}\text{ }|\text{ }\mathbf{x}%
_{i}\mathbf{,}\text{ }z_{i}\mathbf{\in}[\tilde{a},\tilde{b}])\right) \\
&  +\sum_{y_{i}=T_{l,\mathbf{x}}^{\ast}}^{\ }t_{\tilde{a},\mathbf{x}_{i}}%
\ln\left(  p_{\mathbf{\beta,}\alpha}^{\ }(y_{i}\text{ }|\text{ }\mathbf{x}%
_{i}\mathbf{,}\text{ }z_{i}\mathbf{\in}[\tilde{a},\tilde{b}])\right) \\
&  +\sum_{y_{i}=T_{u,\mathbf{x}}^{\ast}}^{\ }(1-t_{\tilde{b},\mathbf{x}_{i}%
})\ln\left(  p_{\mathbf{\beta,}\alpha}^{\ }(y_{i}\text{ }|\text{ }%
\mathbf{x}_{i}\mathbf{,}\text{ }z_{i}\mathbf{\in}[\tilde{a},\tilde
{b}])\right)  \ .
\end{align*}
Then, we define the \textit{modified CML (MCML) estimator} $\mathbf{\hat
{\theta}}_{\text{MCML}}=(\mathbf{\hat{\beta}}_{\text{MCML}},\hat{\alpha
}_{\text{MCML}})$ by
\[
\mathbf{\hat{\theta}}_{\text{MCML}}=\arg\max_{\mathbf{\theta}}\mathcal{L}%
_{\text{MCML}}\mathcal{(}\mathbf{\theta}).
\]

From (\ref{zWeights}) and Theorem \ref{conv a b}, it is easy to show that
\begin{equation}
\ n^{1/2}\left(  W_{\tilde{a},\tilde{b}}(\mathbf{x},\mathbf{\beta,}%
\alpha)-1\right)  =O_{p}(1) \label{Wt1}%
\end{equation}
and therefore%
\begin{equation}
n^{1/2}(p_{\mathbf{\beta,}\alpha}^{\ }(y\text{ }|\text{ }\mathbf{x,}\text{
}Z\mathbf{\in}[\tilde{a},\tilde{b}])-f_{h(\mathbf{x}_{i}^{\text{T}%
}\mathbf{\beta}),\alpha\ }(y))=O_{p}(1). \label{resasin}%
\end{equation}
Then, according to (\ref{resasin}), both $\mathcal{L}_{\text{CML}}%
\mathcal{(}\mathbf{\theta})$ and \ $\mathcal{L}_{\text{MCML}}(\mathbf{\theta
)}$ tend, under the model, to the unconditional likelihood function with rate
$n^{-1/2} $. For this reason we conjecture that both the CML and the MCML
estimator have the same asymptotic distribution than the unconditional
ML\ estimator, that is,
\[
n^{1/2}(\mathbf{\hat{\theta}}_{\text{CML}}-\mathbf{\theta}_{0})\rightarrow
^{D}N_{p}(\mathbf{0},\mathcal{I}^{-1}(\mathbf{\theta}_{0})),
\]
and
\[
n^{1/2}(\mathbf{\hat{\theta}}_{\text{MCML}}-\mathbf{\theta}_{0})\rightarrow
^{D}N_{p}(\mathbf{0},\mathcal{I}^{-1}(\mathbf{\theta}_{0})),
\]
where $\rightarrow^{D}$ denotes convergence in distribution, $N_{p}%
(\mathbf{\mu,\Sigma)}$ the $p$-variate normal distribution with mean
$\mathbf{\mu}$ and covariance matrix $\Sigma$, and $\mathcal{I}(\mathbf{\theta
)}$ the information matrix. This implies that $\mathbf{\hat{\theta}%
}_{\text{CML}}$ and $\mathbf{\hat{\theta}}_{\text{MCML}}$ are both fully
efficient.\medskip

\noindent\textbf{Remark 1}. Empirical results show that, in order to optimize
the finite sample efficiency, with no loss of robustness, it is convenient to
iterate the conditional ML estimator as follows. Given a current value of
$\mathbf{\hat{\theta}}_{\text{CML}}$ (or $\mathbf{\hat{\theta}}_{\text{MCML}}%
$), we compute new RQR-s. Then, we compute new values of $\tilde{a}$ and
$\tilde{b}$ and use them to update $\mathbf{\hat{\theta}}_{\text{CML}}$.
Often, the process converges after a few iterations, but can also move away
from the initial value. In the experiments reported in Section
\emph{\ref{simmod}}, we found that two steps are enough: the efficiency did
not improve using more iterations. Moreover, in the discrete case, the final
estimator slightly depends on the sample $\left\{  u_{1},...,u_{n}\right\}  $
used to compute $\tilde{a}$ and $\tilde{b}$. To remove this dependency, we
propose to average the final step (MCML) over a few replications of this
sample.\medskip

\noindent\textbf{Remark 2}. In certain circumstances, we may use\ a very
simple alternative procedure to compute robust and consistent estimators of
$\beta_{01}$, $\alpha_{0}$, and $\eta_{0}$ in (\ref{simplemodel}). We first
identify a simple model, which is free of the dispersion parameter, and that
can be taken as an approximation of (\ref{simplemodel}). For example, the
Poisson regression model with mean $h(\beta_{01}+\eta_{0}\nu) $ may be taken
as an approximation of the NB model. We then use an available robust procedure
to estimate $\beta_{01}$ and $\eta_{0}$. In the\ NB case, the conditionally
unbiased bounded influence estimators of K\"{u}nsch et al. (1989), implemented
in the R package \textquotedblleft robeth\textquotedblright\ (Marazzi, 1992)
is a natural choice. In the Beta regression case we note that Atkinson (1985)
transforms the response so that the transformed dependent variable (e.g.,
$\log(y/(1-y))$) assumes values on the real line, and then uses it in a linear
regression analysis. Clearly, we may also use a robust regression estimator in
this case, e.g., the MM estimator implemented in the R package
\textquotedblleft robustbase\textquotedblright. Finally, we estimate
$\alpha_{0}$ using (\ref{MEalpha1}). Since the approximate model is not the
correct one, the estimators do not converge to $\beta_{01}$, $\alpha_{0}$, and
$\eta_{0}$. Usual robust estimators converge however to their population
values and can be used to define fixed cut-off values $a$ and\thinspace$b$ for
$Z$, which also converge to their asymptotic values. The CML (MCML) estimator
of $(\beta_{01},\alpha_{0},\eta_{0})$ given $Z\in\lbrack a,b]$ is then
consistent under (\ref{simplemodel}). \medskip

\section{Simulation experiments for NB regression\textsl{\label{simulations}}}

We present simulation results only for the NB regression model (\ref{model}%
)-(\ref{fNB}). We compared the initial estimators $\mathbf{\tilde{\beta}}$ and
$\tilde{\alpha}$ and the final modified CML estimators $\mathbf{\hat{\beta}}$
and $\hat{\alpha}$. In the following, these estimators will be referred as INI
and CML respectively. All cutoff values were adaptive with $\zeta_{1}=0.05$
and $\zeta_{2}=0.95$. In order to compute the MRC estimators we used the
function maxCorGrid of the R package ccaPP (Alfons et al., 2015). The INI
estimator was completed with the help of the weighted MT estimator described
in Appendix \ref{EMT}. In order to estimate $\alpha_{0}$, we used the function
$\psi$ defined by the equation for $\alpha$ of the optimal M estimator M80
described in Marazzi and Yohai (2010, p.174) and available in the
\textquotedblleft robustGLM\textquotedblright\ package. To compute the CML
estimator, we used the standard R optimizer \textquotedblleft
optim\textquotedblright, reparametrizing $\alpha$ with $\sigma=\sqrt{\alpha}$
in order to satisfy the constraint $\alpha>0$. (For a very small number of
contaminated cases, the optimization process diverged; the initial solution
was recorded in such cases.) Only two iterations of the CML procedure were
computed. For comparison, we also computed the GM estimators of Aeberhard et
al. (2014) that will referred as ACH in the following. To compute the ACH
estimator, we used the R function glmrob.nb (available on internet) with the
parameters: bounding.func=`T/T', c.tukey.beta=4, c.tukey.sig=4, as suggested
by the authors, and the option x-weight=hard that provides hard rejection
weights for the covariate observations. We used the following model:%
\begin{align}
&  y\sim F_{\exp(\mathbf{x}^{\text{T}}\mathbf{\beta}_{0}),\alpha_{0}},\text{
\ \ }\mathbf{x=}\binom{1\text{ }}{\mathbf{x}^{\ast}},\text{ \ \ }%
\mathbf{x}^{\ast}\sim N(0,I_{5}),\label{MC1}\\
&  \mathbf{\beta}_{0}=(1.5,0.5,0.25,0,0,0),\text{ \ \ }\alpha_{0}%
=0.8.\nonumber
\end{align}

\subsection{Simulations at the nominal model\emph{\label{simmod}}}

We first performed four experiments with samples of size $n=100$, $400$,
$1000$, $2000$ from (\ref{MC1}) without addition of outliers. For each
experiment, the number of replications was $N=1000$. \ To measure the quality
of an estimator $(\mathbf{\beta,}\alpha\mathbf{)}$ we used the mean absolute
estimation error (MAEE) and the mean absolute prediction error (MAPE). The
MAEE of $\mathbf{\beta}$ is defined by%
\[
\text{MAEE}(\mathbf{\beta})=\frac{1}{N}\sum\limits_{i=1}^{N}\left\Vert
\mathbf{\beta}_{i}^{\#}-\mathbf{\beta}_{0}\right\Vert _{1},
\]
where $\mathbf{\beta}_{i}^{\#}$ is the estimate of $\mathbf{\beta}_{0}$ based
on the $i^{th}$ replication and $||.||_{1}$ denotes the $l_{1}$ norm. The MAEE
of $\alpha$ is defined in a similar way by
\[
\text{MAEE}(\alpha)=\frac{1}{N}\sum\limits_{i=1}^{N}\left\vert \alpha_{i}%
^{\#}-\alpha_{0}\right\vert \text{.}%
\]
The MAPE of the prediction estimator $\mu_{\mathbf{x}}=\exp\left(
\mathbf{x}^{\text{T}}\mathbf{\beta}\right)  $ of $\mu_{0,\mathbf{x}}%
=\exp\left(  \mathbf{x}^{\text{T}}\mathbf{\beta}_{0}\right)  $ is defined as
\[
\text{MAPE}(\mu)=\frac{1}{N}\sum\limits_{i=1}^{N}\left\vert \mu_{i}^{\#}%
-\mu_{0i}\right\vert \text{,}%
\]
where $\mu_{0i}=\exp\left(  \mathbf{x}_{i}^{\#\text{T}}\mathbf{\beta}%
_{0}\right)  $ and $\mu_{i}^{\#}=\exp\left(  \mathbf{x}_{i}^{\#^{\text{T}}%
}\mathbf{\beta}_{i}^{\#}\right)  $ and $\mathbf{x}_{i}^{\#}$ is the $i^{th}$
replication of $\mathbf{x}$. Table 1 reports the empirical relative
efficiencies measured as the ratios of the MAEE and MAPE of the robust
estimators with respect to the corresponding MAEE and MAPE of the ML
estimators.\textsl{\medskip}

\begin{center}
\textit{Table 1. Empirical relative efficiencies of coefficients, dispersion,
and prediction estimators}\medskip
\end{center}

We observe that the relative efficiencies of the initial estimators were low
but could be improved with the help of the final MCML procedure. With the
exception of the dispersion estimator for $n=100$, our final estimator is much
more efficient than the ACH competitor. (The tuning constants of the ACH
estimator were apparently chosen by the authors in order to obtain a
satisfactory degree of robustness.)

\subsection{Simulation with contaminated data}

In another simulation the model (\ref{MC1}) has been contaminated with $10\%$
of pointwise contamination. Preliminary experiments showed that the estimators
were quite sensitive to outlying values of $y$ when $\mathbf{x}^{\ast
}=(3,1,0,0,0)^{\text{T}}$. This value of $\mathbf{x}$ is moderately outlying
with respect to the majority of the covariate observations. Therefore, we used
point contaminations of the form $(\mathbf{x}_{\text{out}}^{\ast
},y_{\text{out}})$ with $\mathbf{x}_{\text{out}}^{\ast}=(3,1,0,0,0)^{\text{T}%
}$ and a response $y_{\text{out}}$ varying in the set $\left\{
0,1,2,10,20,30,40,50,60,70,100,120,180\right\}  $. For each value of
$y_{\text{out}}$, we generated $1000$ samples of size $n=400$ according to
(\ref{MC1}) and then replaced $10\%$ of the observations with identical
outliers of the form $(\mathbf{x}_{\text{out}}^{\ast},y_{\text{out}})$. Table
2\ reports the MAEE and MAPE of the estimators for the different values of
$\ y_{\text{out}}$. (Outliers were excluded in the computation of the MAPE).
The results are also displayed in Figure 1. Both the MAPE and MAE of the
proposed estimators were smaller than those of ACH for most values of
$y_{\text{out}}$.\medskip

\begin{center}
\textit{Table 2. MAEE and MAPE of coefficient, dispersion, and prediction
estimators for varying }$y_{\text{out}}$\textit{.}

\textit{Figure 1. Mean absolute prediction and estimation errors for varying
}$y_{\text{out}}$.
\end{center}

\section{Application to hospital length of stay\textbf{\label{Examples}}}

In modern hospital management, stays are classified into \textquotedblleft
diagnosis related groups\textquotedblright\ (DRGs; Fetter et al., 1980) which
are designed to be as homogeneous as possible with respect to diagnosis,
treatment, and resource consumption. The mean cost of stay of each DRGs is
periodically estimated with the help of administrative data on a national
basis and used to determine \textquotedblleft standard
prices\textquotedblright\ for hospital funding and reimbursement. Typical
stays are reimbursed according to the standard prices, whereas the
reimbursement of exceptional stays (outliers) is subject to special
negotiations among the partners. Since it is difficult to measure cost, length
of stay (LOS) is often used as a proxy. Outliers are usually defined as
observations with a LOS larger that some arbitrary cutoff value. In designing
and refining the groups, the relationship between LOS and other variables
which are usually available on administrative files has to be assessed and
taken into account.\smallskip\ 

We first reconsider the example described in Marazzi and Yohai (2010). In this
example there are not covariables, that is, only the parameters of a NB
distribution are estimated. Table 3 shows the LOS of 32 stays classified into
DRG \textquotedblleft disorders of the nervous system\textquotedblright\ and
we immediately identify three extreme values: $115$, $198$, $374$ days. The
arithmetic means with and without these observations are $25.5$ and $4.4$
days, respectively. We modeled the observed frequencies of LOS$-1$ (note that,
by definition, the minimal LOS is 1) with a NB model. First, we computed the
ML and the \textquotedblleft optimal\textquotedblright\ M estimator referred
as M80 in Marazzi and Yohai (2010). Then, we computed the modified CML
estimator (called CML in the following) with $a=0.05$ and $b=0.95$ based on
two iterations starting from M80, and averaged over $100$ replications of
$\left\{  u_{1},...u_{n}\right\}  $. We also computed the three estimators
(MLE*, M80*, CML*) after removal of the three outliers. The numerical results
are shown in Table 4. They show that M80 and CML provided results which were
similar to MLE* and unaffected by the outliers. The average values of
$\tilde{a}$ and $\tilde{b}$ were $\bar{a}=0.044$ and $\bar{b}=0.953$ from
which we derived $T_{\bar{a}}=1$, $\ T_{\bar{b}}=7$, $t_{\bar{a}}=0.61$, and
\thinspace$t_{\bar{b}}=0.43$. This means that, in the overage, the CML
estimator completely rejected LOS$-1$ values outside the interval $[0,8]$ and
gives weights $0.61$ and $0.57$ to the extremes of this interval.\medskip

\begin{center}
\textit{Table 3. Length of stay of 32 hospital patients.}\medskip

\textit{Table 4. Estimates of LOS-1 mean and LOS-1 dispersion for disorders of
the nervous system.\medskip}\pagebreak
\end{center}

In a second example, we considered a sample of $649$ hospital stays ($256$
male and $393$ female patients) for the \textquotedblleft major diagnostic
category\textquotedblright\ (MDC) \textquotedblleft Diseases and Disorders of
the Endocrine, Nutritional And Metabolic System\textquotedblright. A MDC is
simply a group of DRGs associated with a particular medical specialty. The
data are shown in Figure 2 (two outliers with LOS $=84$ and LOS$\ =122$ fall
beyond the upper limit of the figure).

We studied the relationship between LOS$-1$ and two covariates: Age of the
patient ($x_{1}$ in years) and Sex of the patient ($x_{2}=0$ for males and
$x_{2}=1$ for females). We considered a NB model with exponential link and
linear predictor $\beta_{0}+\beta_{1}x_{1}+\beta_{2}x_{3}+\beta_{3}x_{1}x_{3}%
$. We compared the ML, the ACH, and the complete estimator (called CML in the
following) proposed in Section \ref{Est proc}. The ACH estimator was computed
with the help of the R function glmrob.nb with the tuning parameters suggested
by the authors. The CML step - with $a=0.05$ and $b=0.95$ and two iterations -
was replicated $30$ times with different vectors $\left\{  u_{1}%
,...,u_{n}\right\}  $. The average values of $\tilde{a}$ and $\tilde{b}$ were
$\bar{a}=0.004$ and $\bar{b}=0.973$, from which we derived $T_{\bar
{a},\mathbf{x}_{i}}$, $T_{\bar{b},\mathbf{x}_{i}}$, $t_{\bar{a},\mathbf{x}%
_{i}}$, and $t_{\bar{b}\text{,}x_{i}}$ ($i=1,...,n$). We found that $65$
observations were totally rejected, $62$ fell on the lower limits $T_{\bar
{a},\mathbf{x}_{i}}$ (receiving an average weight $0.96$) and $9$ on the upper
limits $T_{b,x_{i}}$ (with an average weight $0.52$). In Figure 2, the full
outliers are marked by cross signs (x) and the borderline observations by plus
signs (+). Thus, we had about $11\%$ of contamination, mostly located on the
upper tail of the LOS distribution; no leverage point in the covariate space
were present in these data. We also computed the ML estimator (ML*) after
removal of the full outliers. The numerical results are given in Table 5 and
the prediction lines are drawn in Figure 2.\medskip

\begin{center}
\textit{Figure 2. Data: LOS-1 and Age of }$649$\textit{\ patients and fitted
models according to CML and ML. }\medskip

\textit{Table 5. Coefficient (standard errors) and dispersion estimates for
disorders of the endocrine system.}\medskip
\end{center}

We observe that the CML and the ML* coefficient estimates are very close and
quite similar to the ACH estimates. (However, the standard errors provided by
glmrob.nb are surprisingly large.) We also note that the dispersion parameter
is heavily inflated by the contamination. For CML and ML*, the Sex effect
($\beta_{2}$) is significant at the $5\%$ level and the interaction
($\beta_{3}$) is not significant. Instead, for ML, the interaction is
significant at the $5\%$ level, but not the effect of Sex. Thus, the classical
and the robust inferences are different.

Figure 3 shows three uniform qq-plots of randomized tail probabilities\ $z_{1}%
,...,z_{n}$ based on\ different estimates of $\alpha$ and $\beta$. In panel
(a) the ML estimator has been used and the sigmoidal shape suggests that the
estimated model is incorrect. In panel (b), the $z$-values were based on the
modified CML estimator; the plot is more linear but it gradually departs form
the diagonal for increasing quantiles. This suggests that the robustly fitted
model is adequate for a large proportion of data but not for those
corresponding to very large values of $z$. Panel (c) is based on ML* and the
$z$-values corresponding to the full outliers based on CML have been removed
from the plot; this plot follows the diagonal line very well. Finally, the
boxplots in panel (d) compare the distribution of the absolute residuals
(without outliers) based on ML, ACH, CML, and ML*; the two latter ones are
globally smaller than the former ones. We conclude that CML (and ML*) provide
an adequate model for about $90\%$ of the population.\medskip

\begin{center}
\textit{Figure 3. qq-plots of randomized tail probabilities based on ML, CML,
ML with removal of the extreme z-values from the plot, and boxplots of the
absolute residuals of ML, ACH, CML, and ML*.}\bigskip\pagebreak
\end{center}

\section{ Discussion\textbf{\medskip\label{Discussion}}}

In many areas of applied statistics, the data may be affected by a high level
of contamination. An example is the analysis of hospital length of stay, where
contamination levels as high as $10\%$ are not uncommon. For this reason,
different ad hoc rules of trimming had long been used by practitioners to
remove outliers (e.g., Marazzi et al., 1998) from their data. In these
applications, well founded highly robust procedures are needed.

Maronna et al. (1979) showed that classical M and GM estimators of regression
(see e.g., Huber, 1980, Hampel et al., 1986) were unable to combine a high
level of robustness and a high level efficiency: M and GM estimators can be
very efficient, but are very sensitive to outliers in the factor space. This
work stimulated the research on high breakdown-point estimation that provided
LMS, LTS, and S estimators (see e.g., Maronna et al., 2006) just to mention
three among many other procedures. Then, for the usual linear regression
problem, the MM estimators of Yohai (1987) combined high breakdown point and
high efficiency with the help a two step approach: in the first step, a very
robust initial fit (an S estimator) provided the tool for outlier
identification; the second step was based on an efficient estimator (an M
estimator), where the outliers were downweighted. Since then, similar two-step
procedures have been proposed for different models (Marazzi and Yohai, 2004;
Locatelli et al. 2010; Agostinelli et al., 2014).

However, the familiar high breakdown point regression estimators used in the
first step are based on minimization of a robust measure of the residual scale
and, unfortunately, cannot be used for GLMs with a dispersion parameter, such
as NB and Beta regression. The reason is that the residual distribution
depends on the covariates and robust residual measures of scale are not
available in this case. In this paper, we propose a more general approach that
bypasses residual scales.

Our proposal is an original assembly of well known procedures. In the initial
step we use the MRC estimator (Han, 1987) to \ estimate the slopes up to a
scale factor. A very fast algorithm to compute this estimator has recently
been proposed in Alfons et. al (2016). We complete the MRC with the help of a
weighted MT estimator (Valdora and Yohai, 2014) of a simple negative binomial
regression. We then use randomized quantile residuals (Dunn and Smyth, 1996)
to determine adaptive cutoff values $\tilde{a}$ and $\tilde{b}$ using a
procedure similar to the one proposed in Marazzi and Yohai (2004). Influential
outliers are identified by the residuals not belonging to $[\tilde{a}%
,\tilde{b}]$. Finally, we compute a conditional ML, estimator where residuals
belong to $[\tilde{a},\tilde{b}]$. Since, in the absence of outliers,
$\tilde{a}\rightarrow0$ and $\tilde{b}\rightarrow1$, the CML estimator tends
to the ML estimator for $n\rightarrow\infty$. It is therefore fully efficient.

Monte Carlo simulations confirm that our proposal is very efficient under the
model and very robust under point contamination, both in the response and the
covariate distributions. This kind of contamination is unrealistic; however,
it is generally the least favorable one and allows evaluation of the maximal
bias an estimator can incur. The CML estimator for NB regression also resists
to a moderate fraction of excess zeroes in the response. A more vigorous
treatment of this peculiarity of count data should however be approached with
the help of specific models, such as hurdle models (see, e.g. Min and Agresti,
2002, and Cantoni and Zedini, 2009).

We have shown that the proposed method is a useful tool for modelling hospital
length of stay as a function of available covariates, while identifying
influential outliers according to a model based rule. A set of R functions to
compute the proposed estimators is made available as an R package.
\bigskip\pagebreak

{\noindent}{\LARGE Appendices}

\section{Proof of (\ref{zWeights}) \label{proofzw}}

To simplify notations, we just consider the case without covariates; the
extension to the regression case is straightforward. We suppose that
$\mathbf{\vartheta}=(\mu$,$\alpha)$ is given and let $z=F_{\mathbf{\vartheta}%
}(y)-uf_{\mathbf{\vartheta}}(y)$, where $u\sim U[0,1]$. Suppose that $a$ and
$b$ are given cutoff values for $z$ and define, for any $c$,%
\[
y_{\ }^{\ast}(c)=\max\{y:F_{\mathbf{\vartheta}}(y)\leq c\}.
\]
Note that%
\[
F_{\mathbf{\vartheta}}(y^{\ast}(a)+1)-uf_{\mathbf{\vartheta}}(y^{\ast
}(a)+1)\geq a
\]
is equivalent to $\ \smallskip$%
\[
u\leq\frac{F_{\mathbf{\vartheta}}(y^{\ast}(a)+1)-a}{f_{\mathbf{\vartheta}%
}(y^{\ast}(a)+1)}=t_{a}.
\]
Similarly
\[
F_{\mathbf{\vartheta}}(y^{\ast}(b)+1)-uf_{\mathbf{\vartheta}}(y^{\ast
}(b)+1)\leq b
\]
is equivalent to
\[
u\geq\frac{F_{\mathbf{\vartheta}}(y^{\ast}(b)+1)-b}{f_{\mathbf{\vartheta}%
}(y^{\ast}(b)+1)}=t_{b}.
\]
Put $T_{a}=y^{\ast}(a)+2$, $T_{b}=y_{\ }^{\ast}(b)$, and $A=\{y:T_{a}\leq
y\leq T_{b}\}$. We have
\[
\left\{  a\leq z\leq b\right\}  =A\cup\{y=T_{a}-1,\text{ }u\leq t_{a}%
\}\cup\{y=T_{b}+1,\text{ }u\geq t_{b}\},
\]
and then%
\[
P_{\mathbf{\vartheta}}(a\leq z\leq b\text{ }|\text{ }u)=P_{\mathbf{\vartheta}%
}(A)+f_{\mathbf{\vartheta}}(T_{a}-1)I(u\leq t_{a})+f_{\mathbf{\vartheta}%
}(T_{b}+1)I(u\geq t_{b}),
\]
where $P_{\mathbf{\vartheta}}(A)=F_{\mathbf{\vartheta}}(T_{b}%
)-F_{\mathbf{\vartheta}}(T_{a}-1)$. Let $v=I\left(  a\leq z\leq b\right)  $.
Since $E\left[  I(u\leq t_{a})\right]  =P(u\leq t_{a})=t_{a}$, the
distribution of $y$ $|$ $v=1$ is given by
\[
p_{\mathbf{\vartheta}}(y\text{ }|\text{ }v=1)=\left\{
\begin{array}
[c]{ccc}%
\frac{f_{\mathbf{\vartheta}}(y\mathbf{)}}{Q(\mathbf{\vartheta})} & \text{if} &
y\in A,\ \\
\frac{f_{\mathbf{\vartheta}}(y)t_{a}}{Q(\mathbf{\vartheta})} & \text{if} &
y=T_{a}-1,\\
\frac{f_{\mathbf{\vartheta}}(y)(1-t_{b})}{Q(\mathbf{\vartheta})} & \text{if} &
y=T_{b}+1,\\
0 & \text{if} & \text{elsewhere.}%
\end{array}
\right.  \
\]
where
\[
Q(\mathbf{\vartheta})=E\left[  P_{\mathbf{\vartheta}}(a\leq z\leq b\text{
}|\text{ }u)\right]  =P_{\mathbf{\vartheta}}(A)+p_{\mathbf{\vartheta}}%
(T_{a}-1)t_{a}+p_{\mathbf{\vartheta}}(T_{b}+1)(1-t_{b}).
\]
\medskip

\section{Weighted MT estimator of simple regression\label{EMT}}

We describe the use of the weighted MT estimator to compute $\tilde{\beta}%
_{1}^{\ast}(\alpha)$ and $\tilde{\eta}^{\ast}(\alpha)$ introduced in
subsection \ref{inest}. We consider the simple regression model $Y$ $|$ $v\sim
F_{h{({\beta}}_{01}+{\eta}_{0}v{),\alpha}}$. Assuming that $\alpha$ is known,
the weighted MT estimator of $(\beta_{01},{\eta}_{0})$ is defined as follows.%
\begin{equation}
{\ (}\tilde{\beta}_{1}^{\ast}(\alpha),\tilde{\eta}^{\ast}(\alpha))=\arg
\min_{{\ {\beta}}_{0},\beta_{1}}\frac{1}{n}\sum_{i=1}^{n}{w}(x_{i},\hat
{M},\hat{S})\rho(t(y_{i},\alpha)-m(h{({\beta}}_{0}+{\beta}_{1}{x}_{i}%
,\alpha{)),} \label{estm}%
\end{equation}
where $\rho$ is a continuous and bounded function with a unique local minimum
at $0$, $m$ is the function defined by
\begin{equation}
m(\mu,\alpha)=\arg\min_{\gamma}E_{\mu,\alpha}\left(  \rho\left(
t(y,\alpha)-\gamma\right)  \right)  , \label{eq:mdef}%
\end{equation}
$t(y,\alpha)$ is a variance stabilizing transformation and ${w}(x,\hat{M}%
,\hat{S})$ is a nonnegative non-increasing function of $\left\vert x-\hat
{M}\right\vert /\hat{S}$, where $\hat{M}$ and $\hat{S}$ are robust estimators
of location and scale of the covariate $x$. Usually, $\rho$ is taken in the
Tukey's biweight family given by
\[
\rho_{c}^{T}(u)=1-\max\left(  \left(  1-\left(  \frac{u}{c}\right)
^{2}\right)  ^{3},1\right)  .
\]
In our simulations in Section \ref{simulations} and the examples in section
\ref{Examples} with the NB distribution, we used the transformation
\[
t(y,\alpha)=\left\{
\begin{array}
[c]{lll}%
s(y,\alpha) & \text{if} & 0<\alpha<1.3\\
s(y,1.3) & \text{if} & \alpha>1.3
\end{array}
\right.  ,
\]
where
\[
s(y,\alpha)=\sqrt{1/\alpha-0.5}\mathrm{arcsinh}\left(  \sqrt{\frac
{y+3/8}{1/\alpha-3/4}}\right)  \text{.}%
\]
This is a modification of the transformation proposed by Yu (2009) to allow
values of $\alpha$ larger than $4/3$. We take ${w}(x,\hat{M},\hat
{S})=I(\left\vert x-\text{median}(x_{i})\right\vert /$mad$(x_{i})<2)$ and
$h({z})=\exp({z)}$. Since the variance of $t(y,\alpha)$ is almost constant, it
is not necessary to divide the argument of $\rho_{c}$ by a scale estimator.
While the efficiency of the estimator increases with $c$, its degree of
robustness decreases. Since the weighted MT estimator, is used to define an
initial estimator whose efficiency will be improved in further steps, the
value of $c$ is chosen in order to obtain a satisfactory degree of robustness.
By trial and error we obtain the following rule for choosing $c$ as a function
of $\alpha$: $c=1.5\sigma(\alpha)$, where, for each $\alpha,$ $\sigma(\alpha)$
is the constant that approximates the standard deviation of $t(y,\alpha).$ The
value of $\sigma(\alpha)$ is obtained by interpolation the values in the
following Table A1:\medskip

\begin{center}%
\begin{tabular}
[c]{llllllllllllll}\hline
${\small \alpha}$ & {\small .10} & {\small .20} & {\small .30} & {\small .40}
& {\small .50} & {\small .60} & {\small .70} & {\small .80} & {\small .90} &
{\small 1.0} & {\small 1.1} & {\small 1.2} & {\small 1.3}\\\hline
${\small \sigma}$ & {\small .41} & {\small .40} & {\small .39} & {\small .37}
& {\small .36} & {\small .35} & {\small .33} & {\small .32} & {\small .30} &
{\small .29} & {\small .27} & {\small .26} & {\small .24}\\\hline
\end{tabular}
\smallskip\ 

Table A1. Approximated standard deviations of $t(y,\alpha$) for the NB
distribution\medskip
\end{center}

For the Beta distribution, we have Var$_{\mu,\alpha}(y)=\mu(1-\mu)/(1+\alpha)$
and a suitable variance stabilizing transformation (Bartlett, 1947) is given
by%
\[
t(y,\alpha)=\int_{0}^{y}1/\text{Var}_{\mu,\alpha}(y)^{1/2}d\mu=\sqrt{1+\alpha
}\arcsin\left(  \sqrt{y}\right)  \text{.}%
\]
In our experiments we used this transformation for $\alpha\in\lbrack5,50]$ and
link function $h({u})=\exp({u)/(1+}\exp({u)}$. We follow the same approach as
in the\ NB case. The values of the approximated variances can be found in the
following Table A2:\medskip

\begin{center}%
\begin{tabular}
[c]{ccccccccccccccc}\hline
$\alpha$ & {\small 5} & {\small 6} & {\small 7} & {\small 8} & {\small 9} &
{\small 10} & {\small 15} & {\small 20} & {\small 25} & {\small 30} &
{\small 35} & {\small 40} & {\small 45} & {\small 50}\\\hline
$\sigma$ & {\small .42} & {\small .43} & {\small .43} & {\small .44} &
{\small .45} & {\small .45} & {\small .47} & {\small .48} & {\small .48} &
{\small .49} & {\small .49} & {\small .49} & {\small .49} & {\small .49}%
\\\hline
\end{tabular}
\smallskip\ 

Table A2. Approximated standard deviations of $t(y,\alpha)$ for the beta
distribution\medskip
\end{center}

When $\alpha$ is unknown, the estimator $(\tilde{\beta}_{1}^{\ast}%
(\tilde{\alpha}),\tilde{\eta}^{\ast}(\tilde{\alpha}))$ simultaneously
satisfies equations (\ref{estm}) and (\ref{MEalpha1}). To compute an
approximate solution we consider a grid of possible values of $\alpha$, namely
the values in the tables above. For each $\alpha$ in the grid, we first
compute $(\tilde{\beta}_{1}^{\ast}(\alpha),\tilde{\eta}^{\ast}(\alpha))$ and
then the solution $\tilde{\alpha}^{\ast}$ of (\ref{MEalpha1}). The desired
approximation is then defined as the vector $(\tilde{\beta}_{1}^{\ast}%
(\tilde{\alpha}^{\ast}),\tilde{\eta}^{\ast}(\tilde{\alpha}^{\ast}))$ for which
the difference between $\alpha$ and $\tilde{\alpha}^{\ast}$ is minimal.

\section{Proof of Theorem \ref{conv a b}}

\label{proofT1}

We consider the discrete case, where the RQRs are defined by%
\[
z_{i}=F_{h(\mathbf{x}^{\text{T}}\mathbf{\tilde{\beta}}),\tilde{\alpha}}%
(y_{i})-u_{i}f_{h(\mathbf{x}^{\text{T}}\mathbf{\tilde{\beta}}%
),\widetilde{\alpha}}(y_{i}),\ 1\leq i\leq n,
\]
By Assumption B, there exist $A_{0}$ and $B_{0}$ such that, if
\[
D_{n}=\{n^{1/2}|\tilde{\alpha}-\alpha_{0}|\leq A_{0},\text{ }||n^{1/2}%
(\mathbf{\tilde{\beta}}^{\text{T}}-\boldsymbol{\beta}_{0})||\leq B_{0}\},
\]
we have
\[
P(D_{n})\geq1-\zeta_{1}/2\text{.}%
\]
Put
\[
v_{i}=F_{h(\mathbf{x}^{\text{T}}\mathbf{\beta}_{0}),\alpha_{0}}-u_{i}%
f_{h(\mathbf{x}^{\text{T}}\mathbf{\beta}_{0}),\alpha_{0}}(y_{i}),\text{ }1\leq
i\leq n.
\]
Then $\ $the $v_{i}$'s\ $\ $are i.i.d. with distribution $U[0,1].$ By
Assumption C, there exist $K_{1}$ and $K_{2}$ such that%
\[
z_{i}\geq v_{i}-\left(  K_{1}||\boldsymbol{\tilde{\beta}}-\boldsymbol{\beta
}_{0}||+K_{2}|\tilde{\alpha}-\alpha_{0}|\right)
\]
i.e.,%
\[
z_{i}\geq v_{i}-n^{-1/2}B_{n},
\]
where $B_{n}=O_{p}(1)$. Let $e_{0}$ such that, if $M_{n}=\{B_{n}\leq e_{0}\}$,
then
\[
P(M_{n})\geq1-\zeta_{1}/2.
\]
Let $F_{zn}$ and $F_{vn}$ be the empirical distributions of the $z_{i}$'s and
$v_{i}$'s \ respectively. Then, we have%
\begin{equation}
F_{zn}(v)\leq F_{vn}(v+n^{-1/2}B_{n}). \label{FZN}%
\end{equation}
Since
\[
E_{n}=\sup_{v}n^{1/2}|F_{vn}(v)-v|=O_{P}(1),
\]
we get $F_{vn}(v)\leq v+n^{-1/2}E_{n}$. Then, putting $G_{n}=B_{n}+E_{n},$ by
(\ref{FZN}) we obtain%
\begin{equation}
F_{zn}(v)\leq v+n^{-1/2}G_{n}. \label{FZN1}%
\end{equation}
\ In a similar way we get
\begin{equation}
F_{zn}(v)\geq v-n^{-1/2}G_{n}^{\ast}. \label{FZN2}%
\end{equation}
\ where $G_{n}^{\ast}=O_{p}(1)$. Let%
\[
H_{na}(v)=\frac{\sup(F_{zn}(v)-F_{zn}(a),0)}{1-F_{zn}(a)}%
\]
and
\[
A=\{a:\sup_{v\leq\zeta_{1}}(H_{na}(v)-v)\leq0\}.
\]
Then
\[
\tilde{a}=\inf A
\]
Note that $a\in A$ is equivalent to%
\[
F_{zn}(v)\leq v(1-F_{zn}(a))+F_{zn}(a)\text{ for all }v\leq\text{ }\zeta_{1}%
\]
and this is equivalent to
\[
F_{zn}(a)(1-v)\geq F_{zn}(v)-v\text{ for all }v\leq\text{ }\zeta_{1}.
\]
By (\ref{FZN1}) and (\ref{FZN2}) \ a sufficient condition \ for $a\in A$ \ is
that%
\[
(a-n^{-1/2}G_{n}^{\ast})(1-\zeta_{1})\geq n^{-1/2}G_{n}%
\]
or equivalently that $\ $%
\[
a\geq\ n^{-1/2}\left(  \frac{G_{n}}{1-\zeta_{1}}+G_{n}^{\ast}\right)
\]
This implies that
\[
\tilde{a}\leq n^{-1/2}\left(  \frac{G_{n}}{1-\zeta_{1}}+G_{n}^{\ast}\right)
.
\]
proving that $n^{1/2}$ $\tilde{a}$ is bounded in probability. The proof that
$n^{1/2}(\tilde{b}-1)$ is bounded in probability too is similar.\pagebreak

\noindent{\large {References}\medskip}

\noindent Abrevaya J. (1999). Computation of the maximum rank correlation
estimator. Economics Letters, 62, 279--285.\smallskip

\noindent Aeberhard W.H., Cantoni E. and Heritier S. (2014). Robust inference
in the negative binomial regression model with an application to falls data.
Biometrics, 70, 920-931. DOI: 10.1111/biom.12212\smallskip

\noindent Agostinelli C., Marazzi A. and Yohai V.J. (2014). Robust estimators
of the generalized log-gamma distribution. Technometrics, 56(1),
92-101.\smallskip

\noindent Alfons A., Croux C. and Filzmoser P. (2016). Robust maximum
association estimators. Journal of the American Statistical Association. In
press.\smallskip

\noindent Alfons A. (2015). ccaPP: (Robust) canonical correlation analysis via
projection pursuit. R package version 0.3.1, URL
http://CRAN.R-project.org/package=ccaPP.\smallskip

\noindent Amiguet, M. (2011). Adaptively weighted maximum likelihood
estimation of discrete distributions. Ph.D. thesis, Universit\'{e} de
Lausanne, Switzerland.\smallskip

\noindent Austin, P.C., Rothwell, D.M. and Tu, J.V. (2002). A comparison of
statistical modeling strategies for analyzing length of stay after CABG
surgery. Health services \& outcomes research methodology (3), 107-133.
DOI:10.1023/A:1024260023851\smallskip

\noindent Cadigan N. G. and Chen J. (2001). Properties of robust M--estimators
for Poisson and negative binomial data. Journal of Statistical Computation and
Simulation, 70, 273-288.\smallskip

\noindent Cantoni E., and Ronchetti E. (2001). Robust inference for
generalized linear models. Journal of the American Statistical Association,
96(455),1022-1030.\smallskip

\noindent Cantoni E. and Zedini A. (2009). A robust version of the hurdle
model. Cahiers du d\'{e}partement d'\'{e}conom\'{e}trie No 2009.07,
Facult\'{e} des sciences \'{e}conomiques et sociales, \ Universit\'{e} de
Gen\`{e}ve.\smallskip

\noindent Carter E.M. and Potts H.W.W. (2014). Predicting length of stay from
an electronic patient record system: a primary total knee replacement example.
BMC Medical Informatics \& Decision Making 14: 26.
DOI:10.1186/1472-6947-14-26.\smallskip

\noindent{Cribari--Neto F. and Zeiles, A. (2010). Beta regression in R.
Journal of Statistical Software, 34, 1--24.\smallskip}

\noindent Cuesta-Albertos J.A., Matr\'{a}n C. and Mayo-Iscar A (2008).
Trimming and likelihood: robust location and dispersion estimate in the
elliptical model. The Annals of Statistics, 36(5), 2284--2318.\smallskip

\noindent Davison A.C. and Snell E.J. (1991). Residuals and diagnostics. In
Statistical Theory and Modelling: In Honour of Sir David Cox. D.V. Hinkley, N.
Reid and E.J. Snell (editors), 83--106. Chapman and Hall.\smallskip

\noindent Dunn P.K. and Smyth G.K. (1996). Randomized quantile residuals.
Journal of Computational and Graphical Statistics, 5(3), 236-244.\smallskip

\noindent Espinheira P. L., Ferrari S. L. P. and Cribari-Neto, F. (2008).
Influence diagnostics in beta regression. Computational Statistics \& Data
Analysis, 52(9), 4417-4431.\smallskip

\noindent Ferrari S. L. P. and Cribari-Neto F. (2004). Beta regression for
modelling rates and proportions. Journal of Applied Statistics, 31(7),
799--815\smallskip

\noindent Fetter R.B., Shin Y., Freeman J.L., Averill R.F., and Thompson J.D.
(1980). Casemix definition by diagnosis-related groups. Medical care, 18(1),
1-53.\smallskip

\noindent Gervini D. and Yohai V.J. (2002). A class of robust and fully
efficient regression estimators. Annals of Statistics, 30(2),
583-616.\smallskip

\noindent Hampel F.R., Ronchetti E.M., Rousseeuw P.J. and Stahel W.A. (1986).
Robust Statistics: The Approach Based on Influence Functions. Wiley, New
York.\smallskip

\noindent Han A.K. (1987a). Non-parametric analysis of a generalized
regression model: The maximum rank correlation estimator. Journal of
Econometrics, 35(23), 303-316.\smallskip

\noindent Han A.K. (1987b). A non-parametric analysis of transformations.
Journal of Econometrics, 35, (2-3), 191-209.\smallskip

\noindent Hilbe J.M. (2008). Negative Binomial Regression. Cambridge
University press.\smallskip

\noindent Huber P.J. (1980). Robust Statistics. Wiley, New York.\smallskip

\noindent Hunger M., Baumert J. and Holle R. (2011). Analysis of SF-60 index
data: is beta regression appropriate? Value in Health 14, 759-767.\smallskip

\noindent K\"{u}nsch H.R., Stefanski L.A. and Carroll R.J. (1989).
Conditionally unbiased bounded-influence estimation in general regression
models, with applications to generalized linear models. Journal of the
American Statistical Association, 84(406), 460-466.\smallskip

\noindent Locatelli I., Marazzi A. and Yohai V.J. (2010). Robust accelerated
failure time regression. Computational Statistics \& Data Analysis, 55(1),
874-887.\smallskip

\noindent Marazzi A. (1993). Algorithms, Routines, and S-Functions for Robust
Statistics. Wadsworth, Inc., Belmont, California.\smallskip

\noindent Marazzi A., Paccaud F., Ruffieux C. and Beguin C. (1998). Fitting
the distribution of length of stay by parametric models. Medical Care, 36(6),
915-927.\smallskip

\noindent Marazzi A. and Yohai V.J. (2004). Adaptively truncated maximum
likelihood regression with asymmetric errors. Journal of Statistical Planning
and Inference. 122 (1-2), 271-291.\smallskip

{\noindent Marazzi A. and Yohai V.J. (2010). Optimal robust estimates based on
the Hellinger distance. Advances in Data Analysis and Classification.
Springer-Verlag.\smallskip}

\noindent Maronna R.A., Martin R.D. and Yohai V.J. (2006). Robust Statistics
Theory and Methods. Wiley \& Sons, Ltd.\smallskip

\noindent Maronna R., Bustos O. and Yohai V.J. (1979). Bias-and
efficiency-robustness of general M-estimators for regression with random
carriers, Smoothing techniques for curve estimation, 91-116.\smallskip

\noindent Min Y. and Agresti A. (2002). Modeling nonnegative data with
clumping at zero: A survey. Journal of the Iranian Statistical Society,
1,(1-2), 7-33\smallskip

\noindent Nelder J.A. and Wedderburn R. W. M. (1972). Generalized linear
models. Journal of the Royal Statistical Society, Series A, 135 (3),
370-384.\smallskip

\noindent Venables W. N. and Ripley B. D. (1999). Modern Applied Statistics
with S-PLUS. Third edition. Springer.\smallskip

\noindent Rocha, A. V. and Simas, A. B. (2010). Influence diagnostics in a
general class of beta regression models. Test, 20(1), 95-119.\smallskip

\noindent Seow W.J., Pesatori A.C., Dimont E., Farmer P.B., Albetti B, et al.
(2012). Urinary benzene biomarkers and DNA methylation in Bulgarian petrochemi
workers: Study findings and comparison of linear and beta regression models.
PLOS ONE 7: e50471.\smallskip

\noindent Sherman R.P. (1993). The limiting distribution of the maximum rank
correlation estimator. Econometrica, 61(1), 123-137.\smallskip

\noindent Swearingen C.J., Tillley B.C., Adams R.J., Rumboldt Z., Nicholas
J.S., Bandyopadhyay D. and Woolson R.F. (2011). Application of Beta Regression
to Analyze lschemic Stroke Volume in NINDS rt-PA Clinical Trials.
Neuroepidemiology, 37(2), 73-82.\smallskip\ 

\noindent Valdora M. and Yohai V.J. (2014). Robust estimation in generalized
linear models. Journal of Statistical Planning and Inference, 146,
31-48.\smallskip

\noindent Yu G. (2009. Variance stabilizing transformations of Poisson,
binomial and negative binomial distributions. Statistics and Probability
Letters, 79, 1621-1629.\smallskip

\noindent Yohai V.J. (1987). High breakdown-point and high efficiency robust
estimates for regression. Annals of Statistics, 15(2), 642-656.

\pagebreak

\begin{changemargin}{-2cm}{0cm}

\begin{center}
\begin{tabular}{r|ccc|ccc|ccc}
\hline
& \multicolumn{3}{|c}{{\small \ }$\mathbf{\beta }$} & \multicolumn{3}{|c}{%
{\small \ }$\alpha $} & \multicolumn{3}{|c}{${\small \mu }$} \\ \hline
${\normalsize n}$ & {\normalsize INI} & {\normalsize CML} & {\normalsize ACH}
& {\normalsize INI} & {\normalsize CML} & {\normalsize ACH} & {\normalsize %
INI} & {\normalsize CML} & {\normalsize ACH} \\ \hline
{\small 100}  & {\small 0.55} & {\small 0.74} & {\small 0.71} & {\small 0.78} & {\small 0.71} & {\small 0.76} & {\small 0.50} & {\small 0.70} & {\small 0.76} \\ 
{\small 400}  & {\small 0.52} & {\small 0.88} & {\small 0.75} & {\small 0.73} & {\small 0.85} & {\small 0.79} & {\small 0.48} & {\small 0.89} & {\small 0.76} \\ 
{\small 1000} & {\small 0.51} & {\small 0.93} & {\small 0.78} & {\small 0.73} & {\small 0.93} & {\small 0.83} & {\small 0.48} & {\small 0.93} & {\small 0.78} \\ 
{\small 2000} & {\small 0.54} & {\small 0.95} & {\small 0.75} & {\small 0.73} & {\small 0.94} & {\small 0.83} & {\small 0.50} & {\small 0.95} & {\small 0.75} \\ \hline
\end{tabular}%
\medskip 

Table 1. Empirical relative efficiencies of coefficients, dispersion, and prediction estimates.

\vskip1truecm 

\begin{tabular}{c|c|ccccccccccccc}
\hline
${\small y}_{\text{out}}$ &  & {\small 0} & {\small 1} & {\small 2} & {\small 10} & {\small 20} & {\small 30} & {\small 40} & {\small 50} & {\small 60} & {\small 70} & {\small 100} & {\small 120} & {\small 180} \\ 
\hline
& {\small INI} & \multicolumn{1}{|r}
{\small 0.86} & 
\multicolumn{1}{r}{\small 0.57} & \multicolumn{1}{r}{\small 0.51} & 
\multicolumn{1}{r}{\small 0.51} & \multicolumn{1}{r}{\small 0.51} & 
\multicolumn{1}{r}{\small 0.51} & 
\multicolumn{1}{r}{\small 0.51} & \multicolumn{1}{r}{\small 0.51} & 
\multicolumn{1}{r}{\small 0.51} & \multicolumn{1}{r}{\small 0.51} & 
\multicolumn{1}{r}{\small 0.51} & \multicolumn{1}{r}{\small 0.51} & 
\multicolumn{1}{r}{\small 0.51} \\ 
$\mathbf{\beta }$ & {\small CML} & 
\multicolumn{1}{|r}{\small 0.72} & 
\multicolumn{1}{r}{\small 0.76} & \multicolumn{1}{r}{\small 0.78} & 
\multicolumn{1}{r}{\small 0.55} & \multicolumn{1}{r}{\small 0.38} & 
\multicolumn{1}{r}{\small 0.32} & \multicolumn{1}{r}{\small 0.33} & 
\multicolumn{1}{r}{\small 0.37} & \multicolumn{1}{r}{\small 0.42} & 
\multicolumn{1}{r}{\small 0.45} & \multicolumn{1}{r}{\small 0.53} & 
\multicolumn{1}{r}{\small 0.55} & \multicolumn{1}{r}{\small 0.48} \\ 
& {\small ACH} & 
\multicolumn{1}{|r}{\small 1.27}& \multicolumn{1}{r}{\small 1.19} & 
\multicolumn{1}{r}{\small 1.09} & \multicolumn{1}{r}{\small 0.67} & 
\multicolumn{1}{r}{\small 0.46} & \multicolumn{1}{r}{\small 0.41} & 
\multicolumn{1}{r}{\small 0.45} & \multicolumn{1}{r}{\small 0.50} & 
\multicolumn{1}{r}{\small 0.55} & \multicolumn{1}{r}{\small 0.59} & 
\multicolumn{1}{r}{\small 0.70} & \multicolumn{1}{r}{\small 0.75} & 
\multicolumn{1}{r}{\small 0.88} \\ \hline
& {\small INI} & \multicolumn{1}{|r}{\small 0.45} & \multicolumn{1}{r}%
{\small 0.09} & \multicolumn{1}{r}{\small 0.09} & \multicolumn{1}{r}{\small %
0.09} & \multicolumn{1}{r}{\small 0.09} & \multicolumn{1}{r}{\small 0.09} & 
\multicolumn{1}{r}{\small 0.09} & \multicolumn{1}{r}{\small 0.09} & 
\multicolumn{1}{r}{\small 0.09} & \multicolumn{1}{r}{\small 0.09} & 
\multicolumn{1}{r}{\small 0.09} & \multicolumn{1}{r}{\small 0.09} & 
\multicolumn{1}{r}{\small 0.09} \\ 
${\normalsize \alpha }$ 
& {\small CML} & \multicolumn{1}{|r}{\small 0.45} & 
\multicolumn{1}{r}{\small 0.25} & \multicolumn{1}{r}{\small 0.13} & 
\multicolumn{1}{r}{\small 0.18} & \multicolumn{1}{r}{\small 0.23} & 
\multicolumn{1}{r}{\small 0.23} & \multicolumn{1}{r}{\small 0.23} & 
\multicolumn{1}{r}{\small 0.21} & \multicolumn{1}{r}{\small 0.20} & 
\multicolumn{1}{r}{\small 0.19} & \multicolumn{1}{r}{\small 0.15} & 
\multicolumn{1}{r}{\small 0.12} & \multicolumn{1}{r}{\small 0.09} \\ 
& {\small ACH} & \multicolumn{1}{|r}{\small 0.53} & \multicolumn{1}{r}%
{\small 0.12} & \multicolumn{1}{r}{\small 0.10} & \multicolumn{1}{r}{\small %
0.27} & \multicolumn{1}{r}{\small 0.29} & \multicolumn{1}{r}{\small 0.29} & 
\multicolumn{1}{r}{\small 0.27} & \multicolumn{1}{r}{\small 0.26} & 
\multicolumn{1}{r}{\small 0.24} & \multicolumn{1}{r}{\small 0.23} & 
\multicolumn{1}{r}{\small 0.19} & \multicolumn{1}{r}{\small 0.17} & 
\multicolumn{1}{r}{\small 0.12} \\ \hline
& {\small INI} & \multicolumn{1}{|r}{\small 1.92} & \multicolumn{1}{r}%
{\small 1.17} & \multicolumn{1}{r}{\small 1.11} & \multicolumn{1}{r}{\small %
1.11} & \multicolumn{1}{r}{\small 1.11} & \multicolumn{1}{r}{\small 1.11} & 
\multicolumn{1}{r}{\small 1.11} & \multicolumn{1}{r}{\small 0} & 
\multicolumn{1}{r}{\small 1.11} & \multicolumn{1}{r}{\small 1.11} & 
\multicolumn{1}{r}{\small 1.11} & \multicolumn{1}{r}{\small 1.11} & 
\multicolumn{1}{r}{\small 1.11} \\ 
${\normalsize \mu }$ & {\small CML} & \multicolumn{1}{|r}{\small 1.59} & 
\multicolumn{1}{r}{\small 1.91} & \multicolumn{1}{r}{\small 1.69} & 
\multicolumn{1}{r}{\small 1.19} & \multicolumn{1}{r}{\small 0.79} & 
\multicolumn{1}{r}{\small 0.62} & \multicolumn{1}{r}{\small 0.63} & 
\multicolumn{1}{r}{\small 0.74} & \multicolumn{1}{r}{\small 0.88} & 
\multicolumn{1}{r}{\small 1.04} & \multicolumn{1}{r}{\small 1.40} & 
\multicolumn{1}{r}{\small 1.54} & \multicolumn{1}{r}{\small 1.34} \\ 
& {\small ACH} & \multicolumn{1}{|r}{\small 2.68} & \multicolumn{1}{r}%
{\small 2.39} & \multicolumn{1}{r}{\small 2.18} & \multicolumn{1}{r}{\small %
1.40} & \multicolumn{1}{r}{\small 0.97} & \multicolumn{1}{r}{\small 0.78} & 
\multicolumn{1}{r}{\small 0.78} & \multicolumn{1}{r}{\small 0.89} & 
\multicolumn{1}{r}{\small 1.04} & \multicolumn{1}{r}{\small 1.19} & 
\multicolumn{1}{r}{\small 1.64} & \multicolumn{1}{r}{\small 1.91} & 
\multicolumn{1}{r}{\small 2.63} \\ \hline
\end{tabular}%
\medskip 

Table 2. MAEE and MAPE of coefficient, dispersion, and prediction estimates
for varying $y_{\text{out}}$.
\end{center}

\end{changemargin}

\pagebreak

\begin{center}

\begin{tabular}{cccccccccccccc}
\hline
LOS & 1 & 2 & 3 & 4 & 5 & 6 & 7 & 8 & 9 & 16 & 115 & 198 & 374 \\ \hline
frequency & 2 & 6 & 5 & 5 & 4 & 2 & 2 & 1 & 1 & 1 & 1 & 1 & 1 \\ \hline
\end{tabular}
\medskip 

Table 3. Length of stay of 32 hospital patients.
\vskip1truecm

\begin{tabular}{lcccccc}
\hline
          & MLE   & M80  & CML  & MLE* & M80* & CML* \\ \hline
$\mu $    & 24.47 & 3.58 & 3.12 & 3.41 & 3.17 & 3.39 \\ 
$\alpha $ & 3.08  & 0.44 & 0.32 & 0.35 & 0.24 & 0.42 \\ \hline
\end{tabular}
\medskip

Table 4. Estimates of LOS-1 mean and LOS-1 dispersion for disorders of the nervous system.
\vskip1truecm

\begin{tabular}{c|ccccc}
& $\beta _{0}$ & $\beta _{1}$ & $\beta _{2}$ & $\beta _{3}$ & $\alpha $                            \\ 
\hline
ML  & 1.266            & 0.017            & 0.064            & -0.009           & 1.067            \\ 
    & {\small (0.134)} & {\small (0.002)} & {\small (0.178)} & {\small (0.003)} & {\small (0.067)} \\ 
ACH & 1.656            & 0.004            & -1.055           & 0.012            & 0.542            \\ 
    & {\small (0.726)} & {\small (0.011)} & {\small (0.735)} & {\small (0.011)} & {\small (---) } \\ 
CML & 0.899            & 0.017            & -0.269           & -0.002           & 0.593           \\ 
    & {\small (0.113)} & {\small (0.002)} & {\small (0.154)} & {\small (0.003)} & {\small (0.049)} \\ 
ML* & 0.846            & 0.016            & -0.253           & -0.002           & 0.503  \\ 
    & {\small (0.114)} & {\small (0.002)} & {\small (0.156)} & {\small (0.003)} & {\small (0.046)} \\ 
\end{tabular}%
\medskip

Table 5. Coefficient (standard errors) and dispersion estimates for
disorders of the endocrine system.

\end{center}

\ \vskip-2truecm 
\vbox to 18truecm{
\vss\parindent=0pt
\hskip-0.5truecm\hbox{\psfig{figure=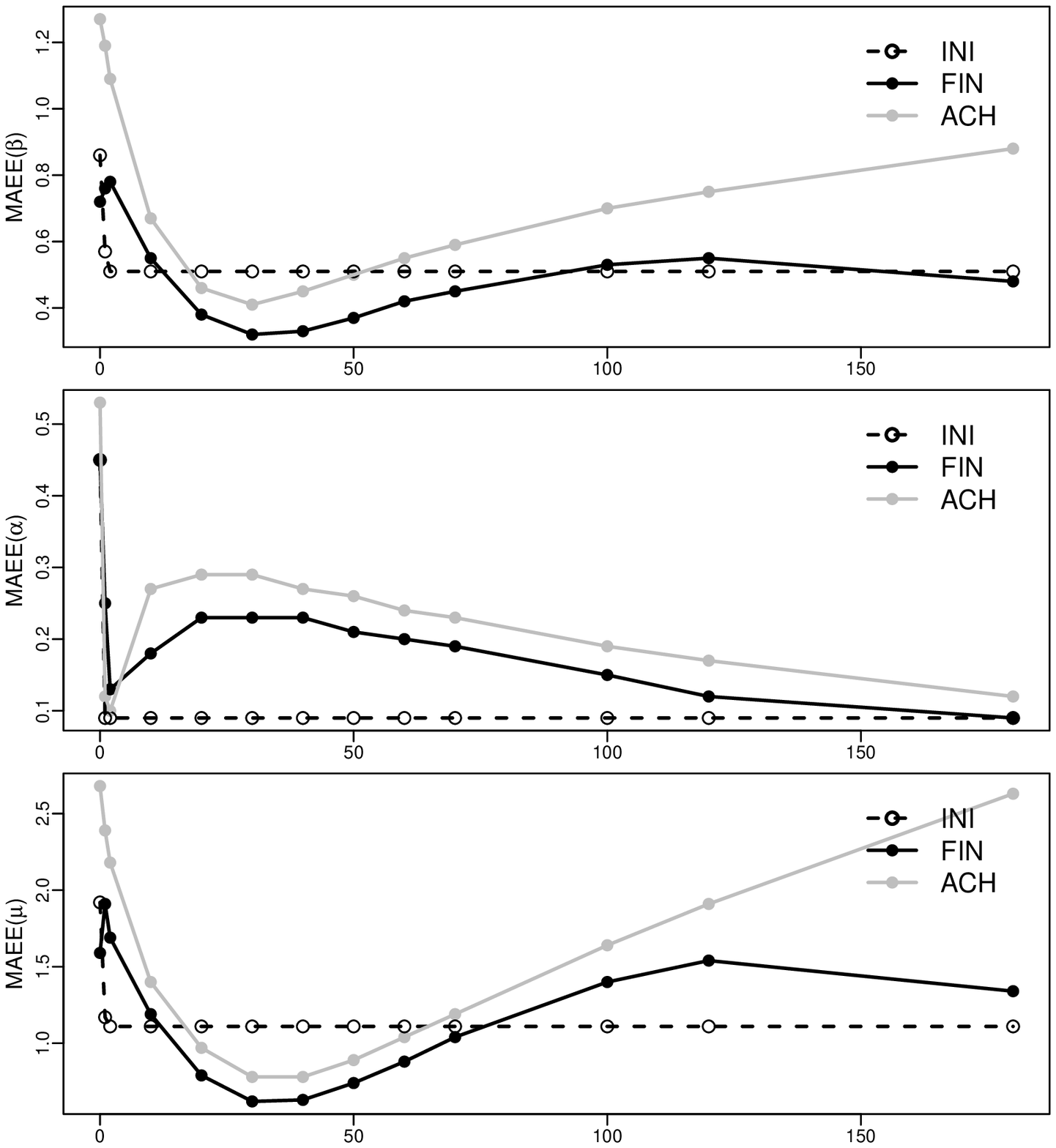,height=16truecm}}} 
\medskip
\hskip-0.5truecm Figure 1. Mean absolute prediction and estimation errors for varying $y_{\text{out}}$.

\pagebreak

%\ \vskip-2truecm 
%\vbox to 16truecm{
%\vss\parindent=0pt
%\hskip-1truecm\hbox{\psfig{figure=M&Y2010.ps,height=16truecm}}} \medskip
%\noindent Figure 2. Observed (black) and estimated LOS frequencies
%according to CML (red) and M80 (blue).

\vfill
\eject

\ \vskip-3truecm 
\vbox to 18truecm{
\vss\parindent=0pt
\hskip-1truecm\hbox{\psfig{figure=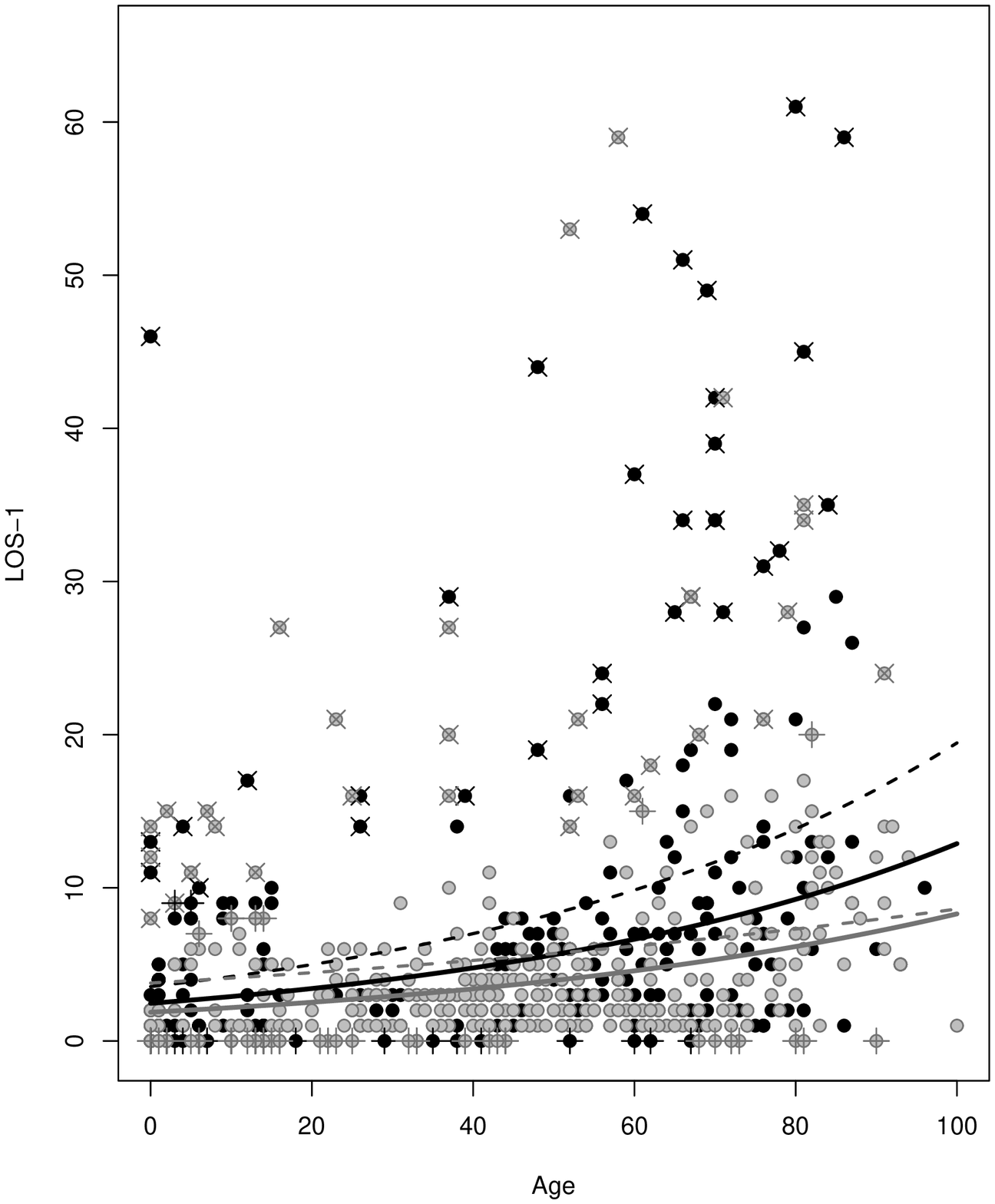,height=18truecm}}} \medskip
\noindent Figure 2. Data: LOS and Age of $649$ patients. 
Black circles are men, gray circles are women. 
Full outliers are marked by cross signs (x); borderline observations by plus signs (+). 
Fitted models according to CML (solid lines) and ML (broken lines): black for men, gray for women

\vfill
\eject

\ \vskip-2truecm 
\vbox to 18truecm{
\vss\parindent=0pt
\hskip-1truecm\hbox{\psfig{figure=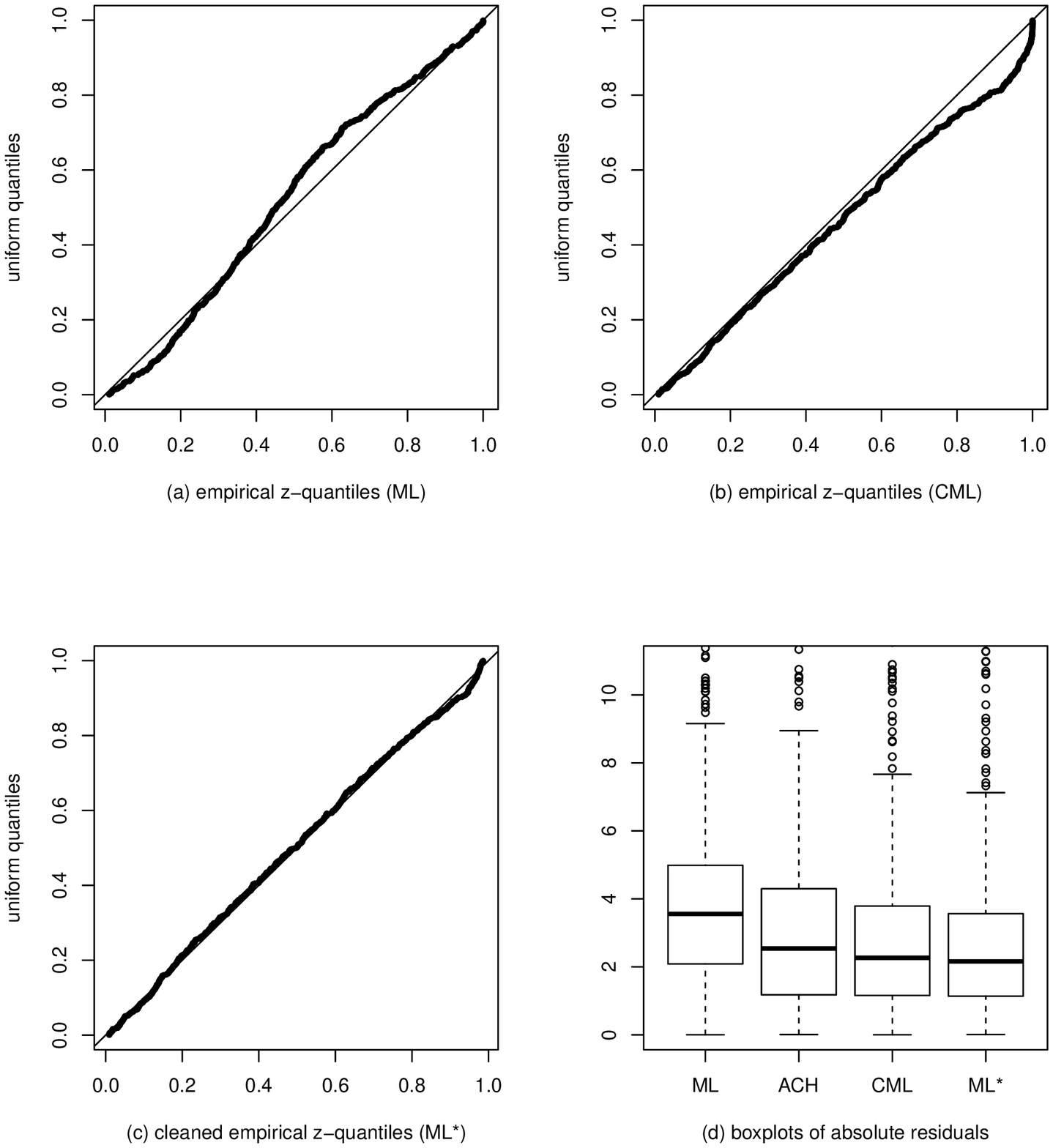,height=18truecm}}} \medskip
\noindent Figure 3. qq-plots of randomized tail probabilities based on: ML
(panel a), CML (panel b), ML with removal of the largest z-values from
the plot (panel c). Panel (d) : boxplots of the absolute residuals of ML, ACH, CML, and ML*.

%\end{changemargin}
\end{document}